\documentclass{article}

\usepackage{amsmath, amssymb} 
\usepackage{bm}
\usepackage{enumerate}
\usepackage[small,nohug,heads=littlevee]{diagrams}
\usepackage{cmll}
\newcommand{\bc}{\begin{C}}
\newcommand{\ec}{\end{C}}
\newcommand{\be}{\begin{equation}}
\newcommand{\ee}{\end{equation}}

\newcommand{\claim}{\begin{Cl}}
\newcommand{\eclaim}{\end{Cl}}
\newcommand{\nb}{\begin{Nb}}
\newcommand{\nbe}{\end{Nb}}
\newcommand{\bl}{\begin{LE}}
\newcommand{\el}{\end{LE}}

\newtheorem{Cl}{Claim}

\newcommand{\bd}{\begin{Def}}
\newcommand{\ed}{\end{Def}}
\newcommand{\bt}{\begin{Th}}
\newcommand{\et}{\end{Th}}

\newtheorem{Th}{Theorem}
\newtheorem{LE}{Lemma}
\newtheorem{C}{Corollary}
\newtheorem{Nb}{Note}
\newtheorem{Def}{Definition}

\begin{document}
 \title{Linear logic in normed cones: probabilistic coherence spaces and beyond
}
\author{Sergey Slavnov
\\  National Research University Higher School of Economics
\\ sslavnov@yandex.ru\\} \maketitle

\begin{abstract}
Ehrhard, Pagani and Tasson \cite{EhrhardPaganiTasson} proposed a model of probabilistic functional programming in a category of normed positive cones and  stable measurable cone maps, which can be seen as a coordinate-free generalization of probabilistic coherence spaces. However, unlike the case of probabilistic coherence spaces, it remained unclear if the model could be refined to a model of classical linear logic.

In this work we consider a somewhat similar category which gives indeed a coordinate-free model of full propositional linear logic with nondegenerate interpretation of additives and sound interpretation of exponentials. Objects are dual pairs of normed cones satisfying certain specific completeness properties, such as existence of norm-bounded monotone weak limits, and morphisms are bounded (adjointable) positive maps.  Norms allow us a distinct interpretation of dual additive connectives as product and coproduct. Exponential connectives are modelled using real analytic functions and distributions that have representations as power series with positive coefficients.

Unlike the familiar case of probabilistic coherence spaces, there is no reference or need for a preferred basis; in this sense the model is invariant. Probabilistic coherence spaces form a full subcategory, whose objects, seen as posets, are lattices. Thus we get a model fitting in the tradition of interpreting linear logic in a linear algebraic setting, which arguably is free from the drawbacks of its predecessors.

Relations with  constructions of \cite{EhrhardPaganiTasson} are left for future research.

\end{abstract}

\section{Introduction}
From the very beginning of linear logic it has been  implicit in notation and terminology that it should be considered as similar to (multi-)linear algebra.

Indeed,  the multiplicative operations,  i.e. tensor, linear implication and linear negation are intuitively interpreted as, respectively, the tensor product of vector spaces, vector space of linear maps and vector space duality; and the additive ones are often understood as corresponding to the product and coproduct of normed spaces. As for the exponentials, they resemble a  ``Fock space''  of power series, i.e. some completion of the free symmetric algebra (as discussed in \cite{BlutePanangadedOldFoundations}, see also \cite{Girard99coherentbanach}).

However, putting this intuition into solid mathematical form turned out to be rather challenging. Different fragments require different structures on vector spaces, and it seems not so easy to put them all together.

On one hand, linear negation requires involutive duality, which is often problematic in infinite dimensions. This leads to considering topological vector spaces and continuous linear maps.

On the other hand, the additive fragment is  better interpreted in the setting of normed vector spaces and norm-bounded linear maps of norm not greater than one, i.e. {\it contractions}. Indeed, in the category of topological vector spaces and continuous linear maps, finite products and coproducts  are isomorphic, and additive connectives get identified. On the other hand,  product and coproduct norms are strictly distinct whenever dimensions are greater than zero. Therefore,  in the category of normed vector spaces and contractions, the two operations are not isomorphic. Unfortunately, duality of normed vector spaces is usually not involutive.

Finally, for  exponentials it is necessary to consider some version of the free symmetric algebra (as in \cite{BlutePanangadedOldFoundations}) and  this  leads to spaces of power series, i.e.  analytic functions. Then it is not always clear how to equip such spaces with norms (since analytic functions globally are unbounded).

In the seminal work \cite{Girard99coherentbanach}  Girard proposed to interpret dual formulas as  {\it dual pairs} of Banach spaces, calling them {\it coherent Banach spaces}.  This gave a perfectly natural, non-degenerate interpretation of the multiplicative-additive fragment, however  failed for the exponentials.

    For the case of exponentials it was proposed to consider the Banach space of functions analytic in the unit ball, but there were issues with the behaviour at the boundary. Unfortunately, functions arising from linear logic proofs tend to map the interior of the unit ball to its closure, and there is no reasonable way to extend a general analytic function from the open ball to its boundary. In particular,  composing these functions was problematic.

Another model was developed  in \cite{Ehrhard_Koethe} in the context of {\it Koethe sequence spaces}. This gives a successful interpretation of full linear logic.  However, since Koethe sequence spaces do not have any norms, there is no distinction between (finite) products and coproducts, and dual additive connectives are identified. Thus, the model of additives is degenerate.
A similar degeneracy occurs in the category of {\it convenient vector spaces} \cite{BluteEhrhardTasson}, which models {\it differential linear logic} (see \cite{diff-lambda}, \cite{diff-nets}), where exponential connectives play a crucial role. 

A more recent model of \cite{Kerjean} is based on a very general class of vector spaces  and interprets full linear logic, but remains additively degenerate in a similar way.

Finally it turned out that the category of {\it probabilistic coherence spaces} (PCS), initiated in \cite{Girard_Logic_quantic} and developed in \cite{DanosErhahrdPCS}, provides a nondegenerate model of the full system, including the additives. Omitting some technicalities, a PCS can be described as a dual pair of ({\it real}) Banach spaces, which are spaces of sequences on the same index set. In a more algebraic language, these are Banach spaces with a fixed basis. Having a basis allows us speaking about {\it positive} (or, rather, non-negative) vectors, represented as non-negative sequences, and positive maps, represented as non-negative matrices. PCS morphisms can be described as bounded positive maps.

 In such a setting,  formulas are interpreted as sequence spaces, and proofs, as positive elements (sequences and matrices). Exponential connectives correspond to spaces of power series with positive coefficients converging in the unit ball (which still have an interpretation as {\it real} analytic functions). Somewhat remarkably, in this case we do not encounter problems with exponentials typical for coherent Banach spaces. A conceptual reason for that is that analytic functions defined in the open unit ball by positive power series have canonical extensions to the boundary (although not necessarily continuous). 

 A major drawback of the PCS model is that it is very non-invariant. All spaces come with  fixed bases. On the other hand, it has eventually become apparent that what makes the model work is not the choice of bases, but the  structure of {\it positivity}, i.e. of a partial order (see \cite{EhrhardPaganiTasson}, \cite{SlavnovSeqSpaces}). In fact, the preferred bases of PCS themselves are determined by the partial order. Sequence spaces, seen as posets (with elementwise ordering) are vector lattices, and preferred bases consist of ``minimal'' positive elements (elements $v$ such that $0\leq u\leq v$ implies $u=\lambda v$ for some $\lambda\leq 1$).

 This suggests
  that the whole construction can be reformulated in an invariant fashion using only the language of partially ordered vector spaces. Or, since eventually we are interested only in positive elements and maps, in the language of {\it abstract positive cones} (in the sense of \cite{Selinger2004}). However it soon becomes evident that the lattice structure, characteristic for PCS, is, in fact, not necessary.

  In particular, in \cite{EhrhardPaganiTasson} a coordinate-free generalization of PCS was proposed for modeling probabilistic functional programming, where more general positive cones are considered. These cones are equipped with norms, similar to vector space norms, and satisfy a variant of {\it completeness}  property: every norm-bounded monotone nondecreasing sequence must have a least upper bound. It turns out that such cones can be organized into a cartesian closed category with specific {\it stable cone maps} as morphisms. (This gives already a model of intuitionistic logic, but in order to interpret probabilistic computation further refinements were proposed, leading to the category of {\it measurable cones} and {\it measurable stable maps}).
  The question of modeling classical linear logic in such a setting remained open.

In this work we propose to consider normed positive cones in dual pairs of partially ordered vector spaces, or, simply, dual pairs of normed cones. Duality alone would allow us to interpret the multiplicative-additive fragment of linear logic along the lines of \cite{Girard99coherentbanach}. But in order to model the exponential fragment we, similarly to \cite{EhrhardPaganiTasson}, introduce certain completeness property. However our version of completeness is different: {\it any norm-bounded monotone sequence has a weak limit}. (It turns out, though, that completeness in our sense implies completeness in the sense of \cite{EhrhardPaganiTasson}.) Here weak limit is understood in the following sense. If $(P,Q)$ is a dual pair of cones, and $\{v_n\}$ is a sequence in $P$, then $v$ is the weak limit of $\{v_n\}$ iff for any $\phi\in Q$ it holds that
$$\langle\phi,v\rangle=\lim_{n\to\infty}\langle \phi,v_n\rangle.$$
Completeness then gives us control over convergence of power series needed for exponentials.

A certain exercise is to define a tensor product of such cones, which  must be complete itself. This leads us to studying possibilities of  cone completion.

  By the end of the day we obtain a well-defined sound model of the exponential fragment in terms of real analytic functions and distributions, similar to the case of PCS. The category of PCS itself is identified as a proper full subcategory. Relations with stable and measurable maps of \cite{EhrhardPaganiTasson} remain to be clarified.

To conclude the introduction, let us comment on some amusing analogies with noncommutative geometry (probably not too strong).

  A standard example of a non-lattice partially ordered vector space is the space of self-adjoint operators on a Hilbert space. In fact, a classical result \cite{Sherman} states that the self-adjoint part of  a $C^*$-algebra is a lattice iff the algebra is commutative. Our passage from vector lattices, which have preferred bases, to general partially ordered vector spaces resembles passage from ``commutative'' spaces  to general noncommutative ``pointless'' spaces, i.e. general algebras.
In particular, probabilistic (``commutative'') coherence spaces have natural representations as subspaces of sequence spaces, and these are commutative algebras. There are however genuinely non-lattice (``noncommutative'') objects, especially coming from spaces of self-adjoint operators on a Hilbert space. Let us mention that for finite dimensions such a noncommutative construction was hinted by Girard a while ago under the name of {\it quantum coherence spaces} \cite{Girard_Logic_quantic}.

\subsection{Plan of the paper}
Our main objects are {\it dual pairs} either of vector spaces or of partially ordered vector spaces or of vector cones. These also can be described in a more intrinsic way as spaces equipped with {\it weak} topology (in spirit of \cite{Kerjean}). We usually prefer the latter representation.

In Section 2 we recall dual pairs of vector spaces and  weak topologies on vector spaces.

In Section 3 we discuss partially ordered vector spaces and vector cones. Next we introduce dual pairs of  partially ordered vector spaces as well as of vector cones. Then we discuss weak topologies in the partially ordered case. For vector spaces, this is a straightforward adaptation from Section 2. For vector cones, some work is needed. We recall the notion of a {\it uniform space} and {\it uniform topology} and then introduce {\it uniform cones} and {\it weak uniform topologies} on cones. We show that vector cones equipped with weak uniform topology are equivalent to dual pairs of cones.

In Section 4 we introduce tensor and cotensor products of cones. In Section 5 we discuss normed cone  and study cone completion.

Finally, in Section 6 we define {\it coherent cones}, which we use to interpret linear logic, and introduce multiplicative-additive operations on them. Section 7 is concerned with  the exponential fragment.

\subsection{Notation and background}
We assume that the reader is familiar with linear logic (see \cite{Girard}, \cite{Girard2} for an introduction), as well as with its categorical interpretation. The standard reference on linear logic and $*$-autonomous categories is \cite{Seely}. For a modern treatment of categorical semantics of linear logic, especially of the exponential fragment, we refer to \cite{Mellies_categorical_semantics}.

We denote the monoidal product in a $*$-autonomous category as $\otimes$ and call it {\it tensor} product, and we denote duality as a star $(.)^{*}$. The dual of tensor product is called {\it cotensor} and denoted as $\parr$, i.e. $A\parr B=(A^*\otimes B^*)^*$. Monoidal unit is denoted ${\bf 1}$, with its dual ${\bf 1}^*$ denoted $\bot$. Product and coproduct are denoted as, respectively, $\times $ and $\oplus$. The internal homs functor is denoted $\multimap$, as usual.

We also assume that the reader is familiar with basic notions of locally convex vector spaces. We use \cite{Shaefer} for reference.  All vector spaces in this paper are real. 

\section{Dual pairs and reflexivity}
The first thing for interpreting linear logic in the context of vector spaces is to get involutive duality. In the finite-dimensional case we have the usual duality of vector spaces. However, when we come to infinite dimensions, there are different ways to generalize this duality,  which, in general, produce non-involutive operations. One of the most direct ways to deal with this problem is to use {\it dual pairs}.

Recall that a {\it dual pair} (also called a {\it dual system} or, simply, a {\it duality})  is a pair $(E,F)$  of vector spaces equipped with a bilinear pairing
$$\langle.,.\rangle:F\times  E\to {\bf R},$$
such that for any $v\in E$ with $v\not=0$ there exists $\phi\in F$ with $\langle \phi,v\rangle\not=0$, and
for any $\phi\in F$ with $\phi\not=0$ there exists $v\in E$ with $\langle \phi,v\rangle\not=0$ (i.e., the pairing is {\it nondegenerate}).

For finite-dimensional $E$ there is only one (up to a natural isomorphism) potential partner to form a dual pair, namely the dual space $E^*$. In the infinite-dimensional case there are many different choices for the dual space, and a dual pair is a way to fix one such choice.

A map of dual pairs $(E,F)$, $(E', F')$ is a pair of linear maps $$L:E\to E',\mbox{ } M:F'\to F$$ satisfying for all $v\in E$, $\phi\in F'$ the {\it adjointness condition}
\be\label{adjointness}
\langle M\phi,v\rangle=\langle \phi,Lv\rangle.
\ee

Obviously, such maps compose, and dual pairs can be organized in a category.

The {\it dual} $(E,F)^*$ of a dual pair $(E,F)$  is the dual pair $(F,E)$ equipped with the same pairing, written with the opposite order of arguments. Obviously, this duality is involutive $(E,F)^{**}=(E,F)$. It also extends to maps in the obvious way: $(L,M)^*=(M,L)$, which gives us a contravariant functor from dual pairs to dual pairs.

\subsection{Weak topologies and reflexivity}
Of course, a dual pair is just a vector space with explicitly specified dual. Always keeping this dual explicit may lead to somewhat cumbersome formulations and expressions (at least, from the notational point of view). We can hide duals from notation by using  topology \cite{Kerjean}.

Indeed, any dual pair $(E,F)$ gives rise to particular locally convex Hausdorff topologies on its members $E$ and $F$.

The {\it$\sigma(E,F)$-weak} (or simply {\it weak}) topology on the space $E$ determined by the dual pair $(E,F)$ is the topology of pointwise convergence on elements of $F$ (where $E$ is identified with a subspace of functions on $F$).

In a greater detail, the {\it$\sigma(E,F)$-weak} topology is defined by saying that a net $\{v_\alpha\}$ in $E$ converges to an element $v\in E$ iff for any $\phi\in F$ the net $\{\langle\phi, v_\alpha\rangle\}$ converges to $\{\langle \phi,v\rangle\}$.

In terms of open sets, the base of zero neighborhoods for the {\it$\sigma(E,F)$-weak} topology consists of the sets
$$U_{\phi_1,\ldots,\phi_k,\epsilon}=\{v\in E|\quad\forall i=1,\ldots,k |\langle\phi_i,v\rangle|<\epsilon\},$$
where $k\in{\bf N}$, $\phi_1,\ldots,\phi_k\in F$ and $\epsilon>0$.

\nb\label{weak topology is l.c. Hausdorff}
For a dual pair  $(E,F)$, the $\sigma(E,F)$-weak topology is locally convex and Hausdorff.
\nbe
{\bf Proof} The topology is Hausdorff because the pairing between $E$ and $F$ is nondegenerate. Local convexity is immediate.
$\Box$
%
\smallskip

In the sequel, unless this leads to  confusion, we will speak loosely about weak topology without mentioning explicitly the dual pair that determines it.

 A crucial  property of weak topology determined by a dual pair $(E,F)$ is the following (see, for example, \cite[IV.1.2]{Shaefer}.
 \bl\label{dual}
For a dual pair $(E,F)$ the space of weakly continuous linear functionals on $E$ is $F$. $\Box$
\el

Now let $V$ be an arbitrary topological vector space (TVS).

We define the {\it dual TVS} $V^*$ as the space of continuous linear functionals on $V$, with the topology of pointwise convergence.  We will understand this  topology on $V^*$ as $\sigma(V^*,V)$-weak and,  accordingly, we will call it  weak. (It  is often called {\it *-weak} in literature though.)

We say that a TVS $V$ is {\it weakly reflexive} if the natural inclusion $$V\to V^{**}$$ is a topological isomorphism. (We add the prefix ``weakly'' to avoid confusion with other notions of reflexivity considered in literature.)

Weak reflexivity of $V$ means  that   topology on $V$ is that of pointwise convergence on elements of $V^*$. In particular, if $(E,F)$ is a dual pair, then the weak topology makes $E$  a reflexive TVS.

The natural notion of TVS map is of course that of a continuous linear map.

Now let $V$, $U$ be TVS.

A linear map $L:V\to U$ is {\it adjointable} iff there exists an {\it adjoint map} $L^*:U^*\to V^*$ defined by the equation
\be\label{adjoint map}
\langle L^*\phi,v\rangle=\langle \phi,Lv\rangle.
\ee

Equation (\ref{adjoint map}) extends duality  of TVS to morphisms thus making it a contravariant functor. In fact, Lemma \ref{dual} implies the following.

\nb\label{adjointable}
A linear map of weakly reflexive TVS is continuous iff it is adjointable.
\nbe
{\bf Proof} Let $V,U$ be weakly reflexive, and $L:V\to U$ be a linear map.

If $L$ is continuous, then for any continuous linear functional $\psi$ on $V$ we have a continuous linear functional $L^*\psi=\psi\circ L$ on $U$. This gives us the adjoint map $L^*:U^*\to V^*$.

Assume that $L$ is adjointable. Let $\{v_\alpha\}$ be a converging net  in $V$ and $v$ be its limit. Then for any $\phi\in U^*$ we have the net $\{\langle \phi,Lv_\alpha\rangle\}=\{\langle L^*\phi,v_\alpha\rangle\}$. Since $V$ is weakly reflexive, its topology is that of pointwise convergence on elements of $V^*$. And since $L^*\phi\in V^*$ and the net $\{v_\alpha\}$ converges to $v$, it follows that the net $\{\langle L^*\phi,v_\alpha\rangle\}$ converges to $\langle L^*\phi, v\rangle=\langle\phi, Lv\rangle$. But $\phi\in U^*$ was arbitrary, so $\{Lv_\alpha\}$ converges pointwise  to $Lv$  on all elements of $U^*$. Hence $L$ preserves converging net limits, i.e. is continuous.
 $\Box$
\smallskip

The above discussion readily gives the following.
\nb\label{refl. TVS=dual pairs}
 The category of weakly reflexive TVS and continuous linear maps is equivalent to the category of dual pairs and dual pair maps under the correspondence
 \be\label{TVS2dual}
 V\mapsto(V,V^*).
 \ee
The equivalence preserves duality.
\nbe
{\bf Proof} Correspondence (\ref{TVS2dual}) is obviously a functor from TVS to dual pairs.

The ``inverse'' functor takes a dual pair $(E,F)$ to the vector space $E$ equipped with the $\sigma(E,F)$-weak topology, and a dual pair map $(L,M)$ to the linear map $L$. In this setting we have, by Lemma \ref{dual}, a natural isomorphism  $E^*\cong F$, hence $E$ becomes weakly reflexive, and  $L$  adjointable, hence continuous, by Note \ref{adjointable}.

Now, composing the two functors in the two possible orders, we get, either the identity
(if we start with a weakly reflexive TVS), or the assignment $(E,F)\mapsto(E,E^*),$ where $E$ is seen as a TVS equipped with the $\sigma(E,F)$-weak topology. The latter is a functor isomorphic to the identity by Lemma \ref{dual}.
$\Box$

\section{Adding positivity}
Our ultimate goal is to interpret the exponential fragment of linear logic, which eventually involves analytic maps given by   power series, and we need to control their convergence, which may be problematic at certain points. A possible solution is to use {\it positive} power series, whose behavior is much simpler. This, in turn, leads us to considering positivity and {\it partial order}.

\subsection{Positivity and  cones}
\subsubsection{Partial order in vector spaces}
Let $E$ be a  vector space.

Recall that a {\it  cone} $P$ in the vector space $V$ is any  subset  satisfying
\begin{itemize}
  \item $$\forall\lambda \in {\bf R}_+ \mbox{ }\lambda P\subseteq P;$$
  \item $$P+P\subseteq P.$$
  \end{itemize}

 A more fancy  way to define  a cone in a vector space is to note that any vector space $V$ is a module over the semi-ring ${\bf R}_+$ and to say that a cone $P$ in $V$ is an ${\bf R}_+$-submodule.

A {\it proper cone} $P$ in $V$ is a cone such that   $$P\cap (-P)=\{0\}.$$

A  proper cone $P$  gives rise to  a partial order on $V$.

Recall that a {\it partially ordered  vector space (POVS)} (see \cite[Chapter V]{Shaefer})  is a vector space $V$ equipped with a partial order $\geq$, such that if $u\geq v$ then
 \begin{itemize}
   \item   $u+w\geq v+w$ for all $w$;
   \item $\lambda u\geq \lambda v$ for all $\lambda\geq 0$.
 \end{itemize}

\nb
For any POVS $V$ the set
\be\label{positive cone}
V_+=\{v\in V|\mbox{ }v\geq 0\}
\ee
 is a proper cone.
Conversely, any   proper cone $V_+$ in $V$ determines a POVS structure  on $V$ by:
$$u\geq v\mbox{ iff }u-v\in V_+.$$
\nbe
{\bf Proof} Exercise or see \cite[Chapter V]{Shaefer}. $\Box$
\smallskip

The set $V_+$ is called the {\it positive cone} of the POVS $V$.
Elements of $V_+$ are called {\it positive}, and elements that are differences of positive elements are called {\it regular}.

A linear map $L$ between POVS is {\it positive}, if $u\geq v$ implies $Lu\geq Lv$. The map $L$ is {\it regular}, if it is the difference of two positive ones.

We will say that a POVS is {\it positively generated} if it is spanned by its positive elements, i.e., if all its elements are regular.

A natural notion of a POVS morphism is that of a  positive map.

 \subsubsection{Cones abstractly}
 Eventually we will be interested only in positive elements and positive maps. Therefore it may be more convenient sometimes to  speak directly about cones without reference to ambient vector spaces (which may be non-unique).
We will consider abstract cones in the sense of \cite{Selinger2004}.

 We say  that an {\it (abstract) cone} is  a module over the semi-ring ${\bf R}_+$,  satisfying the following properties:
 \begin{enumerate}[(i)]
   \item $p+q=p+q'$ iff $q=q'$;
   \item $p+q=0$ iff $p=q=0$.
 \end{enumerate}

 A {\it cone map} is an ${\bf R}_+$-linear map of ${\bf R}_+$-modules.

 Any abstract cone $P$ has an {\it intrinsic partial order} defined by
  \be\label{intrinsic order}
  v\geq u\mbox{ if }\exists u'\in P\mbox{ such that }v=u+u'.
   \ee
   Indeed, the relation in (\ref{intrinsic order}) is obviously reflexive and transitive, and it is easy to deduce from conditions (i), (ii) above that it is antisymmetric in the sense that $u\geq v$, $v\geq u$ iff $u=v$.

   Of course, if we see $P$ as a positive cone in a vector space $V$, then the induced partial order on $V$ extends the intrinsic partial order on $P$.
 \nb\label{cone maps are monotone}
 If $P,Q$ are cones, then  any cone map $L:P\to Q$ is monotone: $u\geq v$ implies $Lu\geq Lv$. $\Box$
 \nbe
\smallskip

 It is rather obvious that for a cone $P$ there is a ``minimal'' embedding in a vector space.

 The {\it enveloping vector space} $EP$ of $P$ is the ${\bf R}_+$-module $P\times  P$ quotiented by the equivalence
 \be\label{envelope quotient}
 (u+p,v+p)\sim(u+p',v+p').
  \ee

  We denote the image of $(u,v)\in P\times P$ in $EP$ as $[u,v]$.

  We define on $EP$ multiplication by $-1$  by $-[u,v]=[v,u]$.

   This makes $EP$  an ${\bf R}$-module, i.e. a vector space. The original cone $P$ embeds in $EP$ by the map $v\mapsto [v,0]$, which makes $EP$ a positively generated POVS.

  \nb\label{EP map}
  Let $P,Q$ be abstract cones. Then any cone map $L:P\to Q$ has unique extension to a positive linear map $EL:EP\to EQ$.
  \nbe
{\bf Proof} Exercise. $\Box$

 \nb\label{EP}
 The category of cones and cone maps is equivalent to the category of positively generated POVS and positive maps. The equivalence is given by the correspondence $$P\mapsto EP.$$
 \nbe
 {\bf Proof} Exercise. $\Box$

\subsubsection{Positivity and dual pairs}
 As in the case of vector spaces, we need also an involutive duality for cones or for POVS, and the most direct way is to mimic the construction of dual pairs.

We adapt the definition of a dual pair to the partially ordered case as follows.

A {\it POVS  dual pair} $(E,F)$ is a dual pair,  whose members are positively generated POVS, such that
\begin{itemize}
\item
 for any $\phi\in F$ it holds that $\phi\in F_+$ iff $\forall v\in E_+\mbox{ }\langle\phi,v\rangle\geq0$;
  \item for any $v\in E$ it holds that $v\in E_+$ iff $\forall \phi\in F_+\mbox{ }\langle\phi,v\rangle\geq0$.
 \end{itemize}

    If $(E,F)$, $(E',F')$ are POVS dual pairs, we say that a dual pair map $(L,M):(E,F)\to(E',F')$ is {\it positive} if the maps $L:E\to E'$, $M:F'\to F$ are positive. (In fact it is sufficient to require positivity of any one of the two maps.) We say that $(L,M)$ is {\it regular} if $L$ and $M$ are regular.

    A {\it POVS dual pair map} is a dual pair map that is positive.

    Since positively generated POVS are essentially just abstract cones it may be reasonable to consider directly dual pairs  of cones.

    A {\it cone  dual pair} is a pair $(P,Q)$ of cones together with  an ${\bf R}_+$-bilinear pairing
$$\langle.,.\rangle:Q\times  P\to {\bf R}_+,$$
such that
\begin{itemize}
\item for any $u,v\in P$ it holds that $v\geq u$ iff $\forall \phi\in Q\mbox{ }\langle\phi,v\rangle\geq\langle\phi,u\rangle$;
\item
 for any $\phi,\psi\in Q$ it holds that $\phi\geq\psi$ iff $\forall v\in P\mbox{ }\langle\phi,v\rangle\geq\langle\psi,v\rangle$.
   \end{itemize}

Note that the above definition implies also that the pairing is nondegenerate, i.e. separates points (since  $v\geq u$ and $u\geq v$ implies $u=v$).
\smallskip

 A {\it map of cone dual pairs} $(P,Q)$, $(P',Q')$ is a pair of cone maps $L:P\to P'$, $M:Q\to Q'$ satisfying adjointness condition (\ref{adjointness}).

\nb\label{positive dual pairs}
The category of POVS dual pairs and POVS dual pair maps is equivalent to the category of cone dual pairs cone dual pair maps.
\nbe
{\bf Proof} follows from Note \ref{EP}. $\Box$
\smallskip

Now, as in the case of ordinary dual pairs, we will try to hide explicit duals from notations and define reflexive POVS and reflexive cones.

\subsection{Topology and partial order}\label{topology and partial order}
Let $X$ be a topological space equipped by with a partial order.

We say that the partial order on $X$ is {\it compatible with topology} if the following property holds:
whenever $x\not\geq y$, there exists a neighborhood $U$ of $x$  such that for all $x'\in U$ it holds that $x'\not\geq y$.

Now let $V$ be a
TVS partially ordered by a proper cone $V_+$.
\nb\label{continuity of order}
The partial order on $V$ is compatible with topology iff the cone $V_+$ is closed.
\nbe
{\bf Proof} Exercise. $\Box$
\smallskip

This observation motivates the following definition.

A {\it partially ordered topological vector space (POTVS)} is a TVS $V$ equipped with a closed proper positive cone $V_+$.

Now let $V$ be a general TVS, and $P$ be a cone in $V$.

 The {\it dual cone} $P^*$ of $P$ is the subset of the topological dual space $V^*$ defined by
$$P^*=\{\phi|\mbox{ }\forall v\in P \mbox{ }\langle\phi,v\rangle\geq 0\}.$$

The {\it bidual cone} $P^{**}$ is the subset of $P$ defined by
$$P^{**}=\{v|\mbox{ }\forall \phi\in P^*\mbox{ } \langle\phi,v\rangle\geq 0\}.$$
{\bf Remark} There is an ambiguity in notation for $P^{**}$, because we could interpret it also as a subset of the double dual space $V^{**}$. In our case this is harmless, because we  always consider weakly reflexive $V$.
\bl\label{bidual}
The bidual $P^{**}$ of a  cone $P$ in the topological vector space $V$  coincides with the $\sigma(V,V^*)$-weak closure of $P$ in $V$.
\el
{\bf Proof} The dual cone $P^*$ can be described as
$$P^*=\{\phi\in V^*|\forall v\in-P\mbox{ }\langle\phi,v\rangle<1\}.$$
Then the statement follows from the Bipolar theorem (see \cite[IV.1.5]{Shaefer} for a formulation). $\Box$
\smallskip

Note that the dual cone $P^*$ is always closed in the $\sigma(V^*,V)$-weak topology.

We define the {\it dual POTVS} $V^*$ of a POTVS $V$ as the space $V^*$ equipped with the $\sigma(V^*,V)$-weak topology and the positive cone $V_+^*$.

Lemma \ref{bidual} implies the following.
\nb\label{reflexive POTVS}
If $V$ is a POTVS which is weakly reflexive as a TVS, then $V^{**}$ is  isomorphic to $V$ both as a POVS and as a TVS.
 $\Box$
\nbe
\smallskip

We will be interested in the more restricted case of weakly reflexive positively generated POTVS.

\subsubsection{Reflexive POTVS}
Let us say that a {\it weakly reflexive  POTVS} is a positively generated POTVS which is weakly reflexive as a TVS and whose dual is positively generated as well.

It follows from the definition that the dual of a weakly reflexive POTVS is itself a weakly reflexive POTVS.

The following properties of weakly reflexive POTVS are direct analogues of the corresponding properties of weakly reflexive TVS.

\nb\label{pos. adjointable}
A positive linear map of weakly reflexive POTVS is continuous iff it is adjointable.
\nbe
{\bf Proof} immediate from Note \ref{EP} and Note \ref{adjointable}. $\Box$

\nb\label{refl. POTVS=dual pairs}
 The category of weakly reflexive POTVS and continuous positive maps is equivalent to the category of POVS dual pairs and POVS dual pair maps under the correspondence
 $$
 V\mapsto(V,V^*).
 $$
The equivalence preserves duality.
\nbe
{\bf Proof} same as Note \ref{refl. TVS=dual pairs}. $\Box$

\subsection{Uniformity and cones}
We want to describe positive cones in POTVS intrinsically. It turns out that embedding in a TVS equips a cone not only with a topology, but also with a structure of a {\it uniform space}.

\subsubsection{Uniform spaces}
Recall that a {\it uniformity} $\mathfrak{U}$ (see \cite{Kelley}) on a set $X$ is a nonempty collection of subsets of $X\times X$ satisfying the following properties:
\begin{enumerate}[(i)]
  \item if $U\in\mathfrak{U}$ then $\Delta_X\subseteq U$, where $\Delta_X$ is the {\it diagonal}, $$\Delta_X=\{(x,x)|x\in X\};$$
  \item if $U\in\mathfrak{U}$ then $U^{-1}\in \mathfrak{U}$;
    \item if $U\in\mathfrak{U}$ then there exists $V\in\mathfrak{U}$ such that $V\circ V\subset U$;
  \item if $U,V\in\mathfrak{U}$ then $U\cap V\in\mathfrak{U}$;
  \item if $U\in\mathfrak{U}$ and $U\subseteq V\subseteq X\times X$ then $V\in\mathfrak{U}$.
\end{enumerate}
Elements of a uniformity are treated in (ii), (iii) as binary relations on $X$. The inverse in (ii) means the inverse (or opposite) relation
$$U^{-1}=\{(x,y)|(y,x)\in U\},$$
and  composition in (iii) is composition of relations.

Elements of uniformity are also called {\it entourages}.

A {\it base} of a uniformity $\mathfrak{U}$ is any subfamily $\mathfrak{B}$ of $\mathfrak{U}$ such that any entourage in $\mathfrak{U}$ contains some element of $\mathfrak{B}$ as a subset. In other words, $\mathfrak{U}$ is the upward closure of $\mathfrak{B}$.

\nb\label{uniformity base}
A family $\mathfrak{B}$ forms a base  for some uniformity if $\mathfrak{B}$ satisfies conditions (i)-(iv) in the definition above.
\nbe
{\bf Proof} Exercise or see \cite[Theorems 5.2, 5.3]{Kelley}. $\Box$
\smallskip

A {\it uniform space} is a space equipped with a uniformity.

A prototypical example of a uniform space is metric space. For any metric space $X$ the collection of sets $U_\epsilon$ of the form
$$U_\epsilon=\{(x,y)|d(x,y)<\epsilon\}$$
for $\epsilon>0$ forms a base of a uniformity.

In particular, property (iii) of the above definition corresponds to the fact that for $\epsilon'<\frac{\epsilon}{2}$ we have $U_{\epsilon'}\circ U_{\epsilon'}\subset U_{\epsilon}$.

Given a uniform space $X$, for any $x\in X$ and entourage $U$ we use the notation
$$U[x]=\{y\in X|(x,y)\in U)\}.$$
The collection $\{U[x]|x\in X\}$ can be thought of as a system of uniform neighborhoods on $X$.

The {\it uniform topology} on a uniform space $X$ is defined by saying that a set $A$ is open iff for any element $x\in A$ there is an entourage $U$ such that the whole uniform neighborhood $U[x]$ is contained in $A$.

It is easy to check that uniform topology is indeed a well-defined topology.

Given two uniform spaces $X,Y$, a function $f:X\to Y$ is {\it uniformly continuous} iff for any entourage $V$ in $Y\times Y$ the counterimage $(f\times f)^{-1}(V)$ is an entourage in $X\times X$.
\nb
Uniformly continuous functions are continuous for the corresponding uniform topologies.
\nbe
{\bf Proof} Exercise or see \cite[Theorem 5.9]{Kelley}. $\Box$
\smallskip

Uniform spaces are closed under subspaces, Cartesian products and quotient spaces.
\smallskip

{\bf Subspaces.}
If $X$ is a uniform space and $Y\subseteq X$ is a subset, then the {\it subspace} or {\it relative} uniformity of $Y$ is defined by the collection of sets $U\cap Y\times Y$, where $U\subseteq X\times X$ is an entourage for $X$.

It is immediate that the corresponding uniform topology on $Y$ is the subspace topology induced by the uniform topology of $X$ and that the inclusion $Y\to X$ is uniformly continuous.
\smallskip

{\bf Products.} If $X$, $Y$ are uniform spaces, then the {\it product uniform space} $X\times Y$ is defined by the {\it product uniformity} whose base is the family of all sets of the form
\be\label{product uniformity}
U_1\ast U_2=\{(x,y,x',y')|(x,x')\in U_1,(y,y')\in U_2\},
\ee
where $U_1$ is some entourage for $X$ and $U_2$ is some entourage for $Y$.
\nb
The uniform topology of a product uniform space $X\times Y$ is the product of uniform topologies of the factors $X,Y$.

The projections $$\pi_1:X\times Y\to X,\quad \pi_2:X\times Y\to Y$$
are uniformly continuous.

A function $f:Z\to X\times Y$, where $Z$ is a uniform space, is uniformly continuous, iff the compositions $\pi_1\circ f$, $\pi_2\circ f$ are uniformly continuous.
\nbe
{\bf Proof} Exercise of see \cite[Theorem 5.10]{Kelley} $\Box$

 It follows from the above that the product of uniform spaces is indeed a Cartesian product (in the categorical sense) in the category of uniform spaces and uniformly continuous functions.
 \smallskip

{\bf Quotients.} Let $X$ be a unform space, $Y$, a set, and $\pi:X\to Y$, a surjective function.
\bt[\cite{Himmelberg}]\label{quotient uniformity}
There exists the largest uniformity on $Y$, called {\bf quotient uniformity}, making $\pi$ uniformly continuous.

Quotient uniformity satisfies the following universal property:
 for any  uniform space $Z$ and  function $f:Y\to Z$, it holds that  $f$ is uniformly continuous iff $f\circ\pi:X\to Z$ is uniformly continuous. $\Box$
\et
\smallskip

In the setting as above, the space $Y$ equipped with the quotient uniformity is called the {\it uniform quotient} (of $X$).

It should be noted that the uniform topology induced by a quotient uniformity in general does {\it not} coincide with the corresponding quotient topology \cite{Himmelberg}.
\nb\label{naive quotinet uniformity}
In the setting as above, assume that the uniformity $\mathfrak{U}$ on $X$ is such that the collection
$$\pi(\mathfrak{U})=\{U| U\subseteq Y\times Y\mbox{ and }(\pi\times\pi)^{-1}(U)\in\mathfrak{U}\} $$
also is a uniformity.

Then the quotient uniformity on $Y$ coincides with $\pi(\mathfrak{U})$.
\nbe
{\bf Proof} Exercise. $\Box$

\subsubsection{TVS and uniformity} For our purposes, an important example of a uniform space is a TVS.

 Given a TVS $E$, we define a uniformity by taking as the base of entourages all sets of the form
 \be\label{TVS uniformity}
 \hat U=\{(u,v)\in E\times E|\mbox{ }u-v\in U\},
 \ee
 where $U$ is a  neighborhood of 0 in $E$.
  \nb
  The system defined by (\ref{TVS uniformity}) is indeed a base for a uniformity.
 \nbe
 {\bf {Proof}} In view of Note \ref{uniformity base} it is sufficient to check properties (i)-(iv) of the definition of a uniformity above.

 The only nontrivial case is property (iii).

 Let $U$ be a neighborhood of 0 in a TVS $E$.

 Since addition is continuous, there exists a neighborhood $V$ of 0 in $E$ such that  $V+V\subset U$. Then for any entourage $\hat U$ of form (\ref{uniformity base}) we have that $\hat V\circ \hat V\subset \hat V$. $\Box$
 \smallskip

 Subsets of form (\ref{TVS uniformity}) have an important property of {\it translation invariance}.

 Let us say that a subset $U$ of $E\times E$, where $E$ is an additive semi-group, is {\it translation invariant} if for any $x\in E$ it holds that $(a,b)\in U$ iff $(a+x,b+x)\in U$.

  \nb\label{TVS uniformity properties}
 With uniformity on a TVS $E$ defined by (\ref{uniformity base}), the uniform topology coincides with original topology.

 Addition and  scalar multiplication and all continuous linear maps are uniformly continuous.

 Translation-invariant entourages form a base of uniformity.
  \nbe
  {\bf Proof} Exercise.
  $\Box$

\subsubsection{Uniform cones}
We say that a {\it uniform cone} $P$ is a cone equipped with a uniformity such that
\begin{enumerate}[(i)]
  \item addition $$P\times P\to P$$ and scalar multiplication
  $${\bf R_+}\times P\to P$$
  are uniformly continuous;
  \item intrinsic partial order (\ref{intrinsic order}) on $P$ is compatible with the uniform topology (see Section \ref{topology and partial order});
    \item translation invariant entourages form a base of the uniformity.
  \end{enumerate}
(In (i), the space ${\bf R}_+$ is considered as a uniform space, with the base of uniformity given by sets
$$\{(a,b)\in {\bf R}_+\times{\bf R}_+|\mbox{ }|a-b|<\epsilon \}$$
where $\epsilon >0$.)
\nb\label{cone in POTVS is topological}
Given a POTVS $E$, the positive cone $E_+$ becomes a uniform cone under the subspace uniformity induced by the uniformity of $E$.
\nbe
{\bf Proof} by Note \ref{TVS uniformity properties} and Note \ref{continuity of order}. $\Box$
\smallskip

We make a couple of observations on relationship between uniformities of a positively generated POTVS and its positive cone.

For a POTVS $E$ and a subset $U\subseteq E\times E$ let us denote $$U_+=U\cap E_+\times E_+.$$

Let $$s:E\times E\to E,\quad (u,v)\mapsto u-v,$$
be the subtraction map.

\nb\label{transl. inv. subsets}
Let $E$ be a positively generated POTVS.

For any translation invariant subset $U\subseteq E\times E$ it holds that
\be\label{TVS uniformity from cone uniformity}
U=s^{-1}(s(U_+)).
\ee
\nbe
{\bf Proof} The set $U$ is translation invariant iff $U=s^{-1}(s(U))$.

But $s(U)=s(U+)$, since for any $(u,v)\in U$, writing
$u=u_+-u_-$ and $v=v_+-v_-$, where $u_+,u_-,v_+,v_-\geq 0$, we have $u-v=(u_++v_-)-(v_++u_-)$.
 
 But $(u_++v_-,v_++u_-)=(u+u_-+v_-,v+u_-+v_-)\in U_+$. $\Box$
\nb\label{TVS uniformity from cone uniformity note}
If $E$ is a positively generated POTVS, then a  subset $U\subseteq E\times E$ is an entourage of $E$ iff $U_+$ is an entourage of $E_+$.
\nbe
{\bf Proof} The only if part is obvious.

Assume that $U_+$ is an entourage of $E_+$. Then $U_+=U_0\cap E_+\times E_+$ for some entourage $U_0$ of $E$. The set $U_0$ contains some translation invariant entourage $V$ of $E$, since translation invariant entourages form a base by Note \ref{TVS uniformity}. Then $V_+\subseteq U_+$. It follows from the preceding note that $V\subseteq U$, hence $U$ is an entourage.
$\Box$

\subsubsection{Enveloping space of a uniform cone}
Now let $P$ be a uniform cone.
%
%
%
 Consider the enveloping POVS $EP$.

 We topologize the enveloping space $EP$ by a uniform topology.

 The uniformity is defined by the base consisting of all sets of the form
 \be\label{enveloping uniformity}
 DU=s^{-1}(\pi(U)),
 \ee
 where $U$ is some  translation invariant entourage of $P$, and
 $$\pi:P\times P\to EP,\quad(u,v)\mapsto[u,v],$$
 is the quotient map.
 \bl
 Sets defined by (\ref{enveloping uniformity}) form a base of uniformity.
 \el
 {\bf Proof}
 By Note \ref{uniformity base} we need to check properties (i)-(iv) in the definition of uniformity.

 We prove property (iii) and leave the rest as an exercise.

 Let $U$ be a translation invariant entourage of $P$.
By uniform continuity of addition, there exists a translation invariant  entourage $V$ of $P$ such that $V+V\subset U$. It is easy to check that $DV\circ DV\subset DU$. $\Box$

      \bl\label{EP is the uniform quotient} In the setting as above, the uniformity of $EP$ is the quotient uniformity induced by $\pi:P\times P\to EP$.
    \el
    {\bf Proof} Denoting the uniformities of $P$ and $EP$, respectively, as $\mathfrak{U}$ and $E\mathfrak{U}$, we will prove that $E\mathfrak{U}=\pi(\mathfrak{U})$ and refer to Note \ref{naive quotinet uniformity}.

    A direct computation shows that, for any two subsets $U_1,U_2\subseteq P\times P$ we have
      $$s(\pi\times \pi(U_1\ast U_2))=\pi(U_1+U_2^{-1}).$$
       If $U_1,U_2$ are translation invariant, then the image $\pi\times\pi(U_1\ast U_2)\subseteq EP\times EP$ is translation invariant as well, hence $$\pi\times\pi(U_1\ast U_2)=s^{-1}(s(\pi\times\pi(U_1\ast U_2))),$$ i.e.
       \be\label{main for EP uniform quotient}
       \pi\times\pi(U_1\ast U_2)=s^{-1}(\pi(U_1+U_2^{-1})).
       \ee

    Now let $U\in\pi(\mathfrak{U})$. Then $U$ contains a subset of the form $\pi\times\pi(U_1\ast U_2)$, where $U_1,U_2$ are translation invariant elements of $\mathfrak{U}$. By (\ref{main for EP uniform quotient}) we have that $U\supseteq s^{-1}(\pi(U_1+U_2^{-1}))\supseteq s^{-1}(\pi(U_1+(0,0))=DU_1$.

    So $U\in E\mathfrak{U}$.

Let $U\in E\mathfrak{U}$. Then $U$ contains a subset of the form $DU_0$, where $U_0$ is some translation invariant element of $\mathfrak{U}$. By uniform continuity of addition there is $V\in \mathfrak{U}$ such that $V+V\subset U_0$. By (\ref{main for EP uniform quotient}) we have: $U\supseteq s^{-1}(\pi(V+V))=\pi\times\pi(V\ast V^{-1})$. So $V\in \pi\mathfrak{U}$. $\Box$
   \bc
Addition and scalar multiplication on $EP$ are uniformly continuous. $\Box$
\ec
\bc\label{uniform map ext}
For topological cones $P,Q$, any uniformly continuous cone map $L:P\to Q$ extends to a unique continuous positive map $EL:EP\to EQ$, where topologies on $EP,EQ$ are defined as above.
\ec
\bl\label{P closed in EP}
The image $i(P)$ of $P$ under the inclusion $$i:P\to EP,\quad v\mapsto[v,0]$$ is closed.
\el
{\bf Proof}
Let $w=[u,v]\in EP$ be in the complement of $i(P)$. Then $u\not\geq v$.

By definitions of  uniform cone and unform topology, there exists an entourage $U_0$ of $P$ such that for any $u'\in U_0[u]$ we have $u'\not\geq v$. Without loss of generality $U_0$ is translation invariant.

Let $U=DU_0$. This is an entourage of $EP$.

Then $U[w]=\{w'\in EP|w-w'\in \pi(U_0)\}$.

 Assume that  $i(P)$ in $EP$ has nonempty intersection with $U[w]$. Then there exists $u'\in P$ such that $w'=[u',0]\in U[w]$, hence
 $w-w'=[u,v+u']\in \pi(U_0)$. But $U_0$ is translation invariant, which means precisely that $U_0=\pi^{-1}(\pi(U_0))$. So $(u,v+u')\in U_0$, and $v+u'\in U_0[u]$. But $v+u'\geq v$, which gives us a contradiction.

Thus $U[w]$ is contained in the complement of $i(P)$, and since $w$ was arbitrary it follows that $i(P)$ is closed.
 $\Box$
 \smallskip

It follows that $EP$ topologized as the uniform quotient of $P\times P$ is a POTVS.

We will say that $EP$ with such a topology is the {\it enveloping POTVS} of $P$.
\nb\label{EV_+=V}
Let $V$ be a positively generated POTVS, and let $V_+$ be its positive cone considered as a uniform cone under the subspace uniformity. Then the enveloping POTVS $EV_+$ and $V$ are isomorphic as POTVS.
\nbe
{\bf Proof} The isomorphism is given by the map $$F:[u,v]\mapsto u-v.$$

It follows from Note \ref{TVS uniformity from cone uniformity note} and Note \ref{transl. inv. subsets} that $F\times F$ and $F^{-1}\times F^{-1}$ preserve entourages (compare (\ref{TVS uniformity from cone uniformity}) and (\ref{enveloping uniformity})). $\Box$
\nb\label{embedding topological cone} The uniformity of a uniform cone $P$ coincides with the subspace uniformity induced by the inclusion in $EP$.
    \nbe
    {\bf Proof}
    Let us treat $P$ as a subset of $EP$.

    For any subset $A\subseteq EP$, we have  $\pi^{-1}(A)=s^{-1}(A)\cap P\times P$. For  any translation invariant subset $B$ of $P$, we have $B=\pi^{-1}(\pi(B))$. Thus for any translation invariant entourage $U$ of $P$ we have $$DU\cap P\times P=s^{-1}(\pi(U))\cap P\times P=\pi^{-1}(\pi(U))=U.$$

    Since translation invariant subsets form a base, the statement is proven. $\Box$
\bc The category of uniform cones and uniformly continuous cone maps is equivalent to the category of positively generated POTVS and positive continuous linear  maps.

The equivalence is given by the assignment $P\mapsto EP$.
\ec
{\bf Proof} follows from Note \ref{EP}, Note \ref{EV_+=V}, Note \ref{embedding topological cone} and Corollary \ref{uniform map ext}. $\Box$

\subsubsection{Duality and reflexivity}
If $P$ is a cone and $Q$ is a collection of cone maps $P\to {\bf R}_+$, we define the {\it uniformity of pointwise convergence} on elements of $Q$ by the  base consisting of subsets of the form
\be\label{dual cone uniformity}
U_{\phi_1,\ldots,\phi_k,\epsilon}=\{(u,v)\in P\times P|\forall i=1,\ldots k|\langle\phi,u\rangle-\langle\phi,v\rangle|<\epsilon\},
\ee
where $k\in{\bf N}$, $\phi_1,\ldots,\phi_k\in Q$ and $\epsilon>0$.

It is easy to check that the above system is indeed a uniformity, and the corresponding uniform topology is the topology of pointwise convergence on elements of $Q$.

If $(P,Q)$ is a cone dual pair, we define the {\it $\sigma(P,Q)$-weak uniformity},  on $P$ as the uniformity of pointwise convergence on elements of $Q$, and the {\it $\sigma(P,Q)$-weak topology}, as the corresponding uniform topology. As usual, we will often say simply weak uniformity (topology) without specifying the dual pair.

We define the  {\it uniform dual}  $P^*$ of a uniform cone  $P$ as the space of uniformly continuous cone maps $P\to{\bf R}_+$ equipped with the $\sigma(P^*,P)$-weak uniformity.

Now, if $P$ is a uniform cone, then every element $v\in P$ defined by pairing a ${\bf R}_+$-linear functional on $P^*$, and  it is immediate from definition (\ref{dual cone uniformity}) of uniformity for $P^*$ that this functional is uniformly continuous.

We say that  a uniform cone $P$
is {\it reflexive}
if the natural map
\be\label{P2P**}
P\to P^{**}
\ee
 is an  isomorphism of uniform spaces.

Reflexive uniform cones correspond to weakly reflexive POTVS.

\nb\label{(EP)*=E(P)*}
If $P$ is a reflexive uniform cone then the enveloping POTVS $EP$ is weakly reflexive with
\be\label{(EP)*=E(P)* eq}
(EP)^*\cong E(P^*).
\ee
\nbe
{\bf Proof}
The pairing between $P$ and $P^*$ extends in the obvious way to a pairing between $EP$ and $E(P^*)$. It is easy to see that the latter is nondegenerate since the former is nondegenerate.
%
%

It follows that $(EP,E(P^*))$ is a dual pair.

Now the uniformity on $P$ is that of pointwise convergence on elements of $P^*$. It follows from formula (\ref{enveloping uniformity}) defining the uniformity on $EP$ that the uniformity on $EP$ is defined by pointwise convergence on elements of $P^*$ as well, hence the topology on $EP$ is that of pointwise convergence on elements of $P^*$.
The latter is equivalent to the $\sigma(EP,E(P^*))$-weak topology, since $E(P^*)$ is the algebraic span of $P^*$. The statement follows from Lemma \ref{dual}.
 $\Box$
\nb\label{cone is refl. POTVS}
If $V$ is a weakly reflexive POTVS, then the cone $V_+$ equipped with the subspace uniformity is  reflexive.
\nbe
{\bf Proof} By Note \ref{EV_+=V}, we have that $V\cong EV_+$. By Corollary \ref{uniform map ext}, we have that the space $V_+^*$ of uniformly continuous cone maps from $V_+$ to ${\bf R}_+$ coincides with the space of positive continuous linear maps from $V$ to ${\bf R}$. The latter set is just the positive cone of the dual POTVS $V^*$. The same analysis applies to $V^*$ and $V_+^*$. I.e., the cone $V_+^{**}$ coincides with the positive cone in the double dual $V^{**}$. The latter cone is just $V_+$ since the POTVS $V$ is assumed weakly reflexive. $\Box$
\smallskip

 The following must be routine by now.
\bc\label{refl.cones=refl.POTVS}
The category of  reflexive uniform cones and uniformly continuous cone maps is equivalent to the category of weakly reflexive POTVS and positive continuous linear maps. The equivalence is geiven by the assignment $P\mapsto EP$. The equivalence preserves duality. $\Box$
\ec
\smallskip

Putting together Note \ref{refl. POTVS=dual pairs} and Corollary \ref{refl.cones=refl.POTVS}, we get the summary.
\nb\label{cones and pairs}
The following categories are equivalent:
\begin{itemize}
\item the category of POVS dual pairs and POVS dual pair maps;
\item the category of cone dual pairs and cone dual pair maps;
\item the category of weakly reflexive POTVS and  positive continuous linear maps;
\item the category of  reflexive uniform cones and uniformly continuous cone maps. $\Box$
\end{itemize}
\nbe
\smallskip
\bc\label{adjointable positive}
A map $L:P\to Q$ of  reflexive uniform cones is uniformly continuous iff there exists an  adjoint map $L^*:Q^*\to P^*$ defined by (\ref{adjoint map}).
\ec
{\bf Proof} Exercise. $\Box$
\bc\label{positive dual}
If $(P,Q)$ is a cone dual pair and $P$ is equipped with the $\sigma(P,Q)$-weak uniformity, then $P^*$ coincides, as a set,  with $Q$.
\ec
{\bf Proof} Exercise. $\Box$

\section{Positivity and tensor product }

Let $(E_i,F_i)$, $i=1,2$, be POVS dual pairs.

Let us say that  a bilinear  functional $$f:E_1\times  E_2\to{\bf R}$$ is {\it separately weakly continuous} if for any $v_i\in E_i$, $i=1,2$, the functionals
$$f(v_1,.):E_2\to{\bf R},\quad f(.,v_2):E_1\to{\bf R}$$
 are weakly continuous.

We say that a bilinear separately weakly continuous functional  $f$ is {\it positive} if for all $v_1\in (E_1)_+$, $v_2\in (E_2)_+$ we have $f(v_2,v_2)\geq 0$. We say that $f$ is {\it regular} if it is a difference of two positive functionals.

Consider the space $F_1\parr_{top} F_2$ of regular bilinear separately weakly continuous functionals on $E_1\times  E_2$.

The cone $(F_1\parr_{top} F_2)_+$ of positive elements makes $F_1\parr_{top} F_2$ a positively generated POVS.

We now specify a dual (i.e. a weak topology) for this space.

Consider the algebraic tensor product $E_1\otimes_{alg} E_2$.

 We have the pairing $\langle.,.\rangle$ between $E_1\otimes_{alg}E_2$ and $F_1\parr_{top} F_2$, given
 for any
 $$w=\sum u_i\otimes v_i\in E_1\otimes_{alg}E_2,$$
  where all $u_i\in E_1$, $v_i\in E_2$, and any functional $f\in F_1\parr_{top} F_2$ by the formula
 \be\label{tensor pairing}
 \langle f,w \rangle=\sum f(u_i, v_i).
 \ee

 Let $E_1\otimes_{top} E_2$ be the space $E_1\otimes_{alg} E_2$ quotiented by the null-space
 $$W=\{w\in E_1\otimes_{alg}E_2|\mbox{ }\langle f,w\rangle=0\forall f\in F_1\parr_{top} F_2\}.$$

 Then pairing (\ref{tensor pairing}) descends to $E_1\otimes_{top} E_2$ and becomes nondegenerate. Thus $(E_1\otimes_{top} E_2,F_1\parr_{top} F_2)$ is a dual pair of vector spaces.

 We define the positive cone $(E_1\otimes_{top} E_2)_+$ in $E_1\otimes_{top}E_2 $ as the dual cone $$(E_1\otimes_{top} E_2)_+=(F_1\parr_{top} F_2)_+^*,$$ which makes $E_1\otimes_{top} E_2$ a POVS.

 Now, the weak topology on $F_1\parr_{top} F_2$ (of pointwise convergence on elements of $E_1\otimes_{top}  E_2$) is just the topology of pointwise convergence on elements of the form $v_1\otimes v_2$, $v_i\in E_i$, $i=1,2$, since they span algebraically the whole $E_1\otimes_{top} E_2$.

 It is immediate then that the positive cone $(F_1\parr_{top} F_2)_+$ is closed in the weak topology, hence $F_1\parr_{top} F_2$, under the weak topology, becomes a POTVS. Then $E_1\otimes_{top} E_2$, equipped with the weak topology,
 is precisely the dual POTVS, and the two POTVS are weakly reflexive.

 Or, using correspondence between weakly reflexive POTVS and POVS dual pairs (Note \ref{refl. POTVS=dual pairs}),  the pair $(E_1\otimes_{top} E_2,F_1\parr_{top} F_2)$ is a POVS dual pair.

In the above setting, we define the {\it topological tensor product} $(E_1,F_1)\otimes_{top}(E_2,F_2)$ of positive dual pairs as the dual pair
$$(E_1,F_1)\otimes_{top}(E_2,F_2)=(E_1\otimes_{top} E_2,F_1\parr_{top} F_2),$$
 and the {\it topological cotensor product} as the dual
 $$(E_1,F_1)\parr_{top}(E_2,F_2)=(E_1\parr_{top} E_2,F_1\otimes_{top} F_2).$$

 Definition for tensor product of weakly reflexive POTVS  is similar. We refer to Note \ref{cones and pairs} and do not fill in the details.

 Now, let $P_1$, $P_2$ be  reflexive uniform cones. Then the enveloping POTVS $EP_1$, $EP_2$ are weakly reflexive and give rise to POVS dual pairs.

 We define the {\it topological tensor product} $P_1\otimes_{top}P_2$ of cones $P_1$, $P_2$ as the cone of positive elements in the topological tensor product of POTVS
 $$P_1\otimes_{top}P_2=(EP_1\otimes_{top} EP_2)_+$$
 equipped with the subspace uniformity
  and
  the {\it topological cotensor product} $P_1\parr_{top}P_2$, as the uniform dual
  $$P_1\parr_{top}P_2=(P_1^*\otimes_{top} P_2^*)^*.$$

  Adaptation to the case of cone dual pairs is similar.

  We leave it to the reader to check that the definitions for cones agree with definitions for POTVS under the correspondence from Note \ref{cones and pairs}.

  We also note that topological cotensor product $P_1\parr_{top} P_2$ can be equivalently described as the cone of
  ${\bf R}_+$-bilinear separately uniformly continuous functionals on $P_1^*\times P_2^*$.

 \subsection{Internal homs}
 Given two POVS dual pairs $A_i=(E_i,F_i)$, $i=1,2$, the space $E_1\multimap E_2$ of regular dual pair maps from $A_1$ to $A_2$ is a vector space in the obvious way. The subset $(E_1\multimap E_2)_+$ of positive maps is a proper generating cone making $E_1\multimap E_2$ a positively generated POVS.

\nb\label{POVS internal hom}
The POVS $E_1\multimap E_2$ and $F_1\parr_{top} E_2$ are isomorphic.
\nbe
{\bf Proof} The isomorphism sends a dual pair map $(L,M):A_1\to A_2$ to the functional
$$(u,\phi)\mapsto\langle\phi, Lu\rangle.$$

In the opposite direction, if $f\in F_1\parr_{top} E_2$, then for any $u\in E_1$ the functional $f(u,.)$ is weakly continuous on $F_2$, hence it is represented as an element of $E_2$. This gives us a linear map from $E_1$ to $E_2$, $u\mapsto f(u,.)$. We also get a linear map from $F_2$ to $F_1$ given by the similar map $\phi\mapsto f(.,\phi)$. This pair of linear maps constitutes a dual pair map from $A_1$ to $A_2$.

That the isomorphism preserves partial order is immediate $\Box$
\smallskip

Using the above isomorphism,
we define the {\it internal homs} dual pair $A_1\multimap A_2$ as
$$A_1\multimap A_2=(E_1\multimap E_2,E_1\otimes_{top} F_2).$$

An adaptation of this definition for weakly reflexive POTVS or cones is left to the reader.


\bt\label{POVS *-aut}
The category of POVS dual pairs and POVS dual pair maps is $*$-autonomous, i.e.
there is a natural bijection $$Hom(A, B\multimap C)\cong Hom(A\otimes_{top} B,C).$$

The same applies to the categories of cone dual pairs, weakly reflexive POTVS and weakly reflexive cones.
\et
{\bf Proof} follows from Note \ref{POVS internal hom} (and Note \ref{cones and pairs}). $\Box$

\section{Adding norms}
It is well agreed that in the setting of vector spaces the additive connectives  of linear  logic correspond to norms, see \cite{Girard99coherentbanach}.

We are now going to consider {\it normed cones}.
\smallskip

%
%

Let $P$ be a cone. We say that a functional
$$||.||:P\to {\bf R}_+$$
is a {\it cone norm} if the following properties hold:
\begin{itemize}
\item  $||\lambda v||=\lambda ||v||$;
\item $||u+v||\leq ||u||+||v||$;
\item $||v||=0$ iff $v=0$;
\item if $u\geq v$ then $||u||\geq||v||$.
\end{itemize}

We define a {\it normed cone} as a cone equipped with a norm.

If $P$ is a normed cone, we say that an ${\bf R}_+$-linear  functional  $\phi:P\to {\bf {R_+}}$ is {\it norm-bounded}, or simply {\it bounded}, iff there exists
\be\label{dual norm}
||\phi||=\sup\limits_{||v||\leq 1}\phi(v)<\infty.
\ee

Similarly, a  cone  map $L$  is bounded iff
\be\label{positive map norm}
||L||=\sup\limits_{||v||\leq 1}||Lv||<\infty.
\ee
A bounded map of norm less or equal to $1$ is a {\it contraction}.

The {\it norm dual cone} $P'$ of the normed cone $P$ is the cone of norm-bounded  ${\bf R}_+$-linear  functionals from  $P$ to ${\bf R}_+$  equipped with {\it dual norm}  (\ref{dual norm}). (We use a prime in the superscript to avoid confusion with the uniform dual.) Note that $P'$ is a well-defined normed cone itself.

\subsection{Normed  dual pairs and reflexive norms}
We say that a {\it normed cone dual pair} is a  cone  dual pair $(P,Q)$, where  $P$,  $Q$ are normed cones, and the pairing is such that for all $\phi\in Q$, $v\in P$ it holds that
\be
||v||=\sup\limits_{\psi\in Q ,||\psi||\leq 1}\langle \psi,v\rangle,\mbox{ }||\phi||=\sup\limits_{u\in P,||u||\leq 1}\langle \phi,u\rangle.
\ee

If  $(L,M):(P,Q)\to (P',Q')$ is a cone dual pair map, then it is immediate that $L$ is norm-bounded in the sense of (\ref{positive map norm}) iff $M$ is, moreover, in this case $||L||=||M||$.

We say that $(L,M)$ is  a {\it map } or a {\it morphism of normed cone dual pairs} if $L$ (hence $M$) is  bounded.

We say that it is a {\it normed cone dual pair contraction} if $L$ is a contraction.

As usual, we define now corresponding reflexive structures.

Let us say that a norm on the uniform cone $P$ is {\it reflexive} if
\begin{itemize}
  \item any uniformly continuous functional in $P^*$ is norm-bounded on $P$, i.e.
$
  P^*\subseteq P',
 $
  \item the {\it canonical injection}
  $  P\to (P^*)'$
    sending $v\in P$ to the functional
  \be\label{P*2P*'}
  \phi\mapsto \langle\phi,v\rangle
  \ee
  is norm-preserving, i.e.
  for any $v\in P$ we have
  \be\label{reflexive norm def}
||v||=\sup\limits_{\phi\in P^* ,||\phi||\leq 1}\langle \phi,v\rangle.
\ee
\end{itemize}

A {\it reflexive uniform normed cone} $P$ is a  reflexive uniform cone equipped with a reflexive norm.
\nb\label{dual reflexive cone}
If $P$ is a reflexive uniform normed cone, then the uniform dual $P^*$ of $P$ equipped with dual norm (\ref{dual norm}) is reflexive itself.
\nbe
{\bf Proof} Exercise. $\Box$
\smallskip

The following must be now routine.
\nb\label{normed pairs and cones}
The category of reflexive uniform normed cones and uniformly continuous contractions is equivalent to the category of normed cone dual pairs and contractions. The equivalence is given by the correspondence $P\mapsto(P,P^*)$ and preserves duality.
\nbe
{\bf Proof} Exercise. $\Box$
\smallskip

We would  also like to have an equivalent category of reflexive normed POTVS. Basically this means an extension of reflexive cone norms to enveloping weakly reflexive spaces that commutes with duality. Unfortunately, we do not know if such an extension is possible in general.

We discuss now certain extension, which indeed commutes with duality in the finite-dimensional case.

\subsection{Norm extensions}
Let $E$ be a POVS equipped with a usual vector space  norm.

%

A norm on $E$ is called {\it regular} if the following properties hold:
\begin{enumerate}[(i)]
  \item for $u,v\in E$, if $-u\leq v\leq u$ then $||v||\leq ||u||$;
  \item for any $v\in E$ and $\epsilon>0$ there exists $u\in E$ such that $u\geq \pm v$ and $||u||<||v||+\epsilon$.
\end{enumerate}

Now let us denote the dual normed space as $E'$. Since $E$ is a POVS, the dual  $E'$  has natural partial ordering defined by the cone of positive functionals.

\nb\label{dual norms coincide}
Let $E$ be a positively generated POVS with a positive cone $P$. Assume that $P$ is equipped with a cone norm, which is extended to a regular norm on the whole $E$. Then for any norm-bounded ${\bf R}_+$-linear functional on $P$, its linear extension to  the whole $E$ is norm-bounded as well. Moreover,  the dual cone norm on $P'$ given by (\ref{dual norm}) coincides with the dual vector space norm on $P'$ inherited from $E'$:
\be\label{dual vector space norm}
||\phi||=\sup\limits_{v\in E,||v||\leq 1}|\langle\phi,v\rangle|.
\ee
\nbe
{\bf Proof} 
The norm on $E$ is regular, so for any $v\in E$ and any $\epsilon>0$ there exists $u\geq\pm v$ with $||u||<||v||+\epsilon$. But, if $u\geq\pm v$, then $u\geq\frac{1}{2}v-\frac{1}{2}v=0$, so $u\in P$.

So, if $\phi\in P'$, then
$$
\sup\limits_{v\in E,||v||\leq 1}|\langle\phi,v\rangle|\leq\sup\limits_{u\in P,||u||<1+\epsilon}\langle\phi,u\rangle
$$
(since  $|\langle\phi,v\rangle|=\max(\langle\phi,v\rangle,\langle\phi,-v\rangle)$ and $\phi$ is a monotone functional.)

Since $\epsilon>0$ is arbitrary the claim is proven. $\Box$

\bt[\cite{Davies}]\label{Davies}  If $E$ is a Banach POVS with a regular norm, then the  norm on $E'$ is also regular.
\et
{\bf Proof} See \cite[Lemma 2.4]{Davies}. $\Box$
\smallskip

Now let $P$ be a normed cone.

We define a norm $||.||_1$ on the enveloping POVS $EP$ by
\be\label{norm ext}
||v||_1=\inf\limits_{u\geq\pm v, u\in P}||u||.
\ee
(This norm is taken from \cite{Davies}.)

\bl\label{ext.norm is regular}
The norm $||.||_1$ above is a well-defined regular vector space norm on $EP$ extending the original cone norm on $P$.
\el
{\bf Proof} Any element $v$ of $EP$ has a representation
$v=v_+-v_-$ with $v_+,v_-\in P$. Then $u=v_++v_-\in P$ and $u\geq\pm v$. This shows that $||v||_1$ is defined for any $v\in EP$.

Also, if $v\in P$ and $u\geq\pm v$, then $u\in P$ and $||u||\geq||v||$. This shows that $||v||_1\geq||v||$. But $v\geq\pm v$, hence $||v||_1\leq||v||$. This shows that $||v||_1=||v||$ for $v\in P$.

Obviously $||\lambda v||_1=|\lambda|\cdot||v||_1$, and $||v||_1=0$ iff $v=0$.

Also, if $u_1\geq\pm v_1$ and $u_2\geq\pm v_2$ with $u_1,u_2\in P$, then $u_1+u_2\geq\pm(v_2+v_2)$ and $||u_1+u_2||\leq||u_1||+||u_2||$. This shows that $||v_1+v_2||_1\leq ||v_1||_1+||v_2||_1$.

This shows that $||.||_1$ is indeed a vector space norm.

Now, if $-u\leq v\leq u$, then $\frac{1}{2}u\geq\pm\frac{1}{2}u$, hence $u=\frac{1}{2}u+\frac{1}{2}u\geq
\frac{1}{2}u-\frac{1}{2}u\geq 0$, so $u\in P$. Also $u\geq\pm v$, hence $||v||_1\leq||u||=||u||_1$. Thus shows that property (i) of regular norm holds.

Property (ii) is obvious from definition. $\Box$
\smallskip

{\bf Remark} Norm extension (\ref{norm ext}) from $P$ to $EP$ is by no means unique. For example, we can define another norm on $EP$ by
$$||v||'=\inf\limits_{v_+,v_-\in P, v=v_+-v_+}(||v_+||+||v_-||),$$ which coincides with (\ref{norm ext}) on $P$, but is different otherwise (as can be seen already from simple two-dimensional examples.)

\subsubsection{In finite dimensions}
\bc\label{fd refl. norm}
If a reflexive uniform cone $P$ is finite-dimensional, than any cone norm on $P$ is reflexive.
\ec
{\bf Proof} The enveloping space $EP$ is finite-dimensional, hence extension (\ref{norm ext}) of the norm on $P$ makes $EP$ a Banach space.

Again, because of finite-dimensionality, topology does not matter, so $P'=P^*$ is just the cone of all positive functionals on $EP$.

Finally, $(EP)'=E(P')$, since $P$ is reflexive.

By Theorem \ref{Davies} and Lemma \ref{ext.norm is regular} the dual vector space norm on $E(P')$ is regular.

By Note \ref{dual norms coincide} the dual vector space norm of $E(P')$ extends the dual cone norm of $P'$.

Applying Note \ref{dual norms coincide} once again we get that the vector space norm of $EP''$ coincides on $P$ with the cone norm of $P''$.

But $EP''\cong EP$ as normed spaces, and the $EP$ norm restricted to $P$ is the original cone norm of $P$. Hence cone norms of $P$ and $P''$ coincide. $\Box$
\smallskip

 Unfortunately we do not know under which conditions the above arguments can generalize to infinite dimensions and we do not have any classification of reflexive cone norms.

For possible generalizations it might be worth noting that there are some stronger versions of  Theorem \ref{Davies},  see \cite{NG}.


\subsection{Completeness}\label{completeness}
Unlike the case of vector spaces, a normed cone does not have any intrinsic metric; a metric would require extending the norm to the whole enveloping space. Thus there is no analogue of Banach (i.e. metric complete) spaces in the setting of normed cones. Yet there is a specific notion of completeness, which will be crucial for our purposes.

Let $P$ be a  reflexive uniform normed cone.  We say that $P$  is a {\it complete},  if any norm-bounded monotone non-decreasing sequence $v_0\leq\ldots\leq v_n\leq\ldots$
in $P$ has a weak limit in $P$.

\subsubsection{Alternative notions of completeness}
It would be more accurate, but also more cumbersome, to use some term like {\it sequential topological order-completeness} rather than just completeness.

In fact there are other notions of completeness for normed cones.

In \cite{EhrhardPaganiTasson}, an alternative definition is proposed, which does not involve any topology, but only  norms and intrinsic partial order. A cone $P$ is {\it complete in the sense of \cite{EhrhardPaganiTasson}} if any norm-bounded non-decreasing sequence in $P$  has a least upper bound. (This could be called {\it sequential order-completeness} without the word ``topological''.)

 It is not hard to see that completeness in our sense implies completeness in the sense of \cite{EhrhardPaganiTasson}, but not vice versa.
\nb\label{topological order-completeness2order completeness}
If a reflexive uniform normed cone $P$ is  complete (in the sense of this work), then for any  norm-bounded monotone non-decreasing sequence $\{v_n\}\subset P$ its weak limit $\lim\limits_{n\to\infty}v_n$ is also its least upper bound in $P$.
\nbe
{\bf Proof} By definition there exists the weak limit $v=\lim\limits_{n\to\infty}v_n\in P$.

Then for any $\phi\in P^*$ we have $\langle \phi,v\rangle=\lim\limits_{n\to\infty} \langle \phi,v_n\rangle$. But the numerical sequence $\{ \langle \phi,v_n\rangle\}$ is monotone non-decreasing (by Note \ref{cone maps are monotone}), hence its limit equals its supremum, so $\langle \phi,v\rangle=\sup\limits_{n} \langle \phi,v_n\rangle$.
It follows from Lemma  \ref{bidual} that the difference $v-v_n\in EP$ belongs to the  closure of $P$ in $EP$.  But, by Lemma \ref{embedding topological cone}, the image of $P$  in $EP$ is closed. Hence $v\geq v_n$ for all $n$.

Now if $v'\in P$ is some other upper bound of $\{v_n\}$, then for any $\phi\in P^*$ it holds that $\langle \phi,v'\rangle\geq\sup\limits_n\langle \phi,v_n\rangle=\langle \phi,v\rangle$. Hence $v'\geq v$. Thus $v=\sup\limits_n v_n$. $\Box$
\smallskip

Let us give an example of a reflexive uniform normed cone which is complete in our sense, but not in the sense of \cite{EhrhardPaganiTasson}.

Consider the Banach space $l^\infty$ of all real sequences $a=\{a_n\}$ bounded in the norm
$$||a||^\infty=\sup\limits_n|a_n|.$$
The space $l^\infty$ is partially ordered by the cone $l_+^\infty$ consisting of non-negative sequences.

Restriction of the norm $||.||^\infty$ makes $l_+^\infty$ a normed cone, and it is immediate that any norm-bounded sequence $\{a^n\}$ of elements of $l_+^\infty$ has the least upper bound $a\in l_+^\infty$ defined by $a_k=\sup\limits_n a_k^n$.
\bt[\cite{Banach}]
There exists a norm-bounded linear functional $LIM:l^\infty\to{\bf R}$ such that
\begin{enumerate}[(i)]
  \item $LIM (a)=\lim\limits_{n\to\infty}a_n$ whenever the limit in the righthand side exists;
  \item $\lim\inf\limits_{n\to\infty}a_n\leq LIM (a)\leq\lim\sup\limits_{n\to\infty}a_n$. $\Box$
\end{enumerate}
\et
(See \cite[7.2.1]{EinsiedlerWard}  for a modern proof in English.)
\smallskip

Condition (ii) above implies that $LIM$ is positive.

Let $(l_+^\infty)'$ be the cone of all positive norm-bounded  linear functionals on $l^\infty$ equipped with the dual norm.

Then $(l_+^\infty)'$ contains the cone $l_+^1$ of nonnegative sequences $b=\{b_n\}$ bounded in the norm
$$||b||^1=\sum b_n,$$ if we define the pairing by
$$\langle b,a\rangle=\sum a_nb_n.$$

For any $a\in l_+^\infty$ we have
$$||a||^\infty=\sup\limits_{b\in l_+^1, ||b||^1\leq 1}\langle b,a\rangle,$$
and it follows that
$$||a||^\infty=\sup\limits_{b\in (l_+^\infty)', ||b||\leq 1}\langle b,a\rangle.$$

Thus $(l_+^\infty,(l_+^\infty)')$ is a normed cone dual pair. 

We equip $l_+^\infty$ with the $\sigma(l_+^\infty,(l_+^\infty)')$-weak uniformity. This makes the topological normed cone $l_+^\infty$ a reflexive normed cone (since normed cone dual pairs correspond to reflexive normed cones).

Consider the sequence $\{e^{\leq n}\}^\infty_{n=0}\subset l_+^\infty$, where $e^{\leq n}$ is defined by
\[ e^{\leq n}_k= \left\{ \begin{array}{ll}
         1 & \mbox{if $k \leq n$};\\
        0 & \mbox{if $k>n$}.\end{array} \right. \]
This sequence is norm-bounded and has  the elelment $e=(1,1,1,\ldots)\in l^\infty_+$ as the least upper bound.

Assume that $l_+^\infty$ equipped with the $\sigma(l_+^\infty,(l_+^\infty)')$-weak topology is complete (in the sense of this work). Then, by Note \ref{topological order-completeness2order completeness}, the  limit of $\{e^{\leq n}\}$ in the $\sigma(l_+^\infty,(l_+^\infty)')$-weak topology is $e$. Since $LIM\in (l_+^\infty)'$, it must be that the sequence $\{LIM (e^{\leq n})\}$ converges to $LIM (e)$ as $n\to\infty$.

But, for all $n$, we have $LIM (e^{\leq n})=0$, and $LIM (e)=1$. We get a contradiction.
\smallskip

It should be noted though that the example above is very non-constructive and relies heavily on the Axiom of choice (for proving existence of $LIM$).

We would expect that in all ``natural'', constructive examples the two notions of completeness coincide.

\subsubsection{Completing reflexive cones}
Our main use of positivity is in dealing with power series when modeling the exponential fragment. Completeness then gives control on their convergence.

However, prior to interpreting exponentials, we need a multiplicative structure; i.e. we will need  tensor products. Since topological tensor products, in general, will  not be complete, some procedure of cone completion  will be crucial.

It turns out that reflexive normed cones have canonical completions.

Below, we will use the following notation.

If $P$
 is a uniform cone equipped with a norm, then $\widetilde P=(P^*)'$ is the cone of all norm-bounded (not necessarily continuous) cone maps $P^*\to{\bf R}_+$, where the uniform dual $P^*$ of $P$ is equipped with the dual norm given by (\ref{dual norm}). The cone $\widetilde P$, in its norm is equipped with the norm dual to that of $P^*$.

 We also use the following terminology.

For a uniform cone $P$, we say that a subspace $P_0$ is a {\it subcone}  of $P$ if $P_0$ is a ${\bf R}_+$-submodule of $P$ closed in the uniform topology.

It is easy to see that a subcone is itself a uniform cone, when equipped with the subspace uniformity.

When the ambient uniform cone $P$ is equipped with a norm, any subcone of $P$ becomes a normed uniform cone under the norm inherited from $P$.

 Now let $P$ be a  reflexive uniform normed cone.

 Let  the dual uniform cone  $P^*$ be equipped with dual norm (\ref{dual norm}). Recall that $P^*$ is reflexive as well (Note \ref{dual reflexive cone}).

 \nb\label{tilde P,P*} The normed cones $\widetilde P $ and $P^*$ form a normed cone dual pair.
 \nbe
 {\bf Proof}
 Since the injection $P\to (P^*)'=\widetilde P $ is norm preserving, we have that for any $v\in P^*$
 $$||v||=\sup\limits_{u\in P,||u||\leq 1}\langle u,v \rangle\leq \sup\limits_{u\in \widetilde P , ||u||\leq 1}\langle u,v\rangle.$$
 On the other hand, obviously,
 $$||v||\geq\sup\limits_{u\in \widetilde P , ||u||\leq 1}\langle u,v\rangle.$$ So
 $$||v||=\sup\limits_{u\in \widetilde P , ||u||\leq 1}\langle u,v\rangle.$$
 Analogous formula for  $u\in \widetilde P $ is obvious. $\Box$
\smallskip

 We equip $\widetilde P $ with the $(\widetilde P ,P^*)$-weak uniformity and the corresponding weak topology. This makes $\widetilde P $ a reflexive uniform normed cone by Note \ref{normed pairs and cones}.

 \nb The normed uniform cone $\widetilde P $ is complete.
 \nbe
 {\bf Proof} If $\{u_n\}$ is a monotone non-decreasing sequence of elements of $\widetilde P $ bounded in norm by some $A\in{\bf R}$, then for any $v\in P^*$ the sequence $\{\langle u_n,v\rangle\}$ is monotone non-decreasing (by Note \ref{cone maps are monotone}) and bounded by $A||v||$, hence it has a limit in ${\bf R}$. Then the functional $u$ defined by
 $\langle u,v\rangle=\lim\limits_{n\to\infty}\langle u_n,v\rangle$ is norm-bounded by $A$, hence $u\in \widetilde P $. But $u$ is precisely the weak limit of $\{u_n\}$. $\Box$
\nb
If $Q$ is a subcone  of $\widetilde P$ containing $P$, then $Q$, equipped with the subspace uniformity and restriction of the norm, becomes  a reflexive uniform normed cone whose dual coincides, as a set, with $P^*$.
\nbe
{\bf Proof} Since $Q$ contains $P$, the pairing of $Q$ and $P^*$ is nondegenerate, and $(Q,P^*)$ is a cone dual pair.

By the same reasoning as in the proof of Note \ref{tilde P,P*} we see that $(Q,P^*)$ is also a normed cone dual pair.

 On the other hand, the subspace topology of $Q$ induced by inclusion in $\widetilde P$ is precisely the $\sigma(Q,P^*)$-weak topology. The statement follows from  Note \ref{normed pairs and cones}. $\Box$
\smallskip

Now consider the  set $\mathfrak{P}$ of complete subcones of $\widetilde P$ containing $P$.
The set $\mathfrak{P}$ is closed under intersection. It follows that there exists the smallest complete subcone $\overline{P}$ of $\widetilde P $ containing $P$. We call $\overline{P}$ the {\it completion} of $P$.

The preceding note implies that the dual of $\overline{P}$ coincides, as a set, with $P^*$.
\bt\label{universal property} With notation as above, the completion $\overline{P}$ satisfies the following universal property.

For any  uniformly continuous cone map $$L:P\to X,$$
  where $X$ is a complete reflexive uniform normed cone, there is a unique uniformly  continuous
  $$\overline{ L}:\overline{P}\to X$$
  with $||\overline{ L}||=||L||$, making the following diagram commute.
 \begin{diagram}
 \overline P&\rTo^{\overline L}&X\\
 \uTo^{i}&\ruTo_{L}&\\
 P&&
  \end{diagram}
Here
$i:P\to \overline P$ is the inclusion map.
\et
{\bf Proof}
 Let $L:P\to X$ be as in the formulation.

 There is  a {\it double adjoint} map $\widetilde L:\widetilde P \to \widetilde X $ defined by
  $$\langle \phi,\widetilde Lv\rangle=\langle L^*\phi,v\rangle,$$
  where $\phi\in X^*$, $v\in \widetilde P $.
  The map $\widetilde L$ is itself adjointable (with the adjoint $L^*$), hence uniformly continuous by Note \ref{adjointable positive}. It is also bounded with $||\widetilde L||=||L^*||=||L||$.

   Since $X$ is reflexive, it is identified  as a subcone of $\widetilde X$. Let $R={\widetilde L}^{-1}(X)$.

  If $\{a_n\}$ is a norm-bounded  non-decreasing sequence in $R$, then $\{\widetilde La_n\}$ is a  norm-bounded non-decreasing (by Note \ref{cone maps are monotone}) sequence  in $X$. Since $X$ is complete, we have that $x=\lim\limits_{n\to\infty} \widetilde L a_n\in X$. Since $\widetilde L$ is continuous, it takes $a=\lim\limits_{n\to\infty}a_n$ to $x$. Hence $a\in R$.

  It follows that the  cone $R$ is complete, hence $\overline P\subseteq R$. Then the restriction of $\widetilde L$ to $\overline P$ is a well defined map to $X$. Its restriction to $P$ coincides with $L$. $\Box$

 \bt\label{completion is complete}
 With notation as above, if $P^*$ is complete in the $\sigma(P^*,P)$-weak topology, then it remains complete in  the $\sigma(P^*,\overline{P})$-weak topology.
 \et
 {\bf Proof} It is enough to show that  $\sigma(P^*,P)$-weak convergence and  $\sigma(P^*,\overline{P})$-weak convergence, when restricted to monotone sequences, coincide on $P^*$.

 Let $R$ be the subcone of elements  $v\in (P^*)'=\widetilde P $ satisfying the following property:
 \begin{itemize}
   \item for any  monotone  non-decreasing sequence $\{\phi_n\}$ in $P^*$ with $\sigma(P^*,P)$-weak limit $\phi\in P^*$ it holds that
       $$\langle\phi,v\rangle =\lim\limits_{n\to\infty}\langle\phi_n,v\rangle.$$
 \end{itemize}

 We are going to show that $R$ is a complete subcone of $\widetilde P $.

So let $\{v_n\}$ be a norm-bounded monotone non-decreasing sequence in $R$. Then there exists $v=\lim\limits_{n\to\infty}v_n\in \widetilde P $.

Let $\{\phi_m\}$ be a monotone non-decreasing sequence in $P^*$ with $\sigma(P^*,P)$-weak limit $\phi\in P^*$. Note that all $\phi_m$ and $\phi$ are uniformly continuous on $\widetilde P $, hence on $R$.

We have
$$
\langle\phi,v\rangle=
\lim\limits_{n\to\infty} \langle\phi,v_n\rangle =$$
$$=\sup_n \langle\phi,v_n\rangle=\sup_n\lim\limits_{m\to\infty}\langle\phi_m,v_n\rangle=$$
$$=\sup_n\sup_m\langle\phi_m,v_n\rangle=
\sup_m\sup_n\langle\phi_m,v_n\rangle=$$
$$\sup_m\lim\limits_{n\to\infty}\langle\phi_m,v_n\rangle=
\sup_m\langle\phi_m,v\rangle=$$
$$=\lim\limits_{m\to\infty}\langle\phi_m,v\rangle.$$


Thus $v\in R$, hence $R$ is complete.

But $P\subseteq R$, hence $\overline{P}\subseteq R$ as well. And this is precisely what we needed to show. $\Box$

The moral of the above is that
if $(P,Q)$ is a normed positive dual pair where  $Q$ is complete in the corresponding weak topology, then it has a completion $(\overline P,Q)$ to a normed positive dual pair, whose both members are complete. The completion is minimal and canonical as is seen from the universal property of Theorem \ref{universal property}.

\section{Coherent cones.}
We are ready to introduce our category for modeling linear logic.

\bd { Coherent cone} is defined by one of the following equivalent structures:
\begin{itemize}
  \item A complete reflexive uniform normed cone $P$ whose dual $P^*$ is also complete;
  \item a normed cone  dual pair  $(P,Q)$ such that $P$, $Q$ are complete for respective weak topologies.
\end{itemize}
\ed

Unless otherwise stated, we will use the first representation and understand coherent cones as complete reflexive unform normed cones.

Morphisms of coherent cones are uniformly continuous contractions.

Since contractions compose, and identities are contractions, it follows that coherent cones can be organized into a category, which we denote as ${\bf CCones}$.

It follows from the very definition that the duality $(.)^*$ equips ${\bf CCones}$ with a contravariant functorial involution.


Now let us make sure that coherent cones indeed exist and show some examples.

\subsection{Examples}\label{coherent cone examples}
\subsubsection{Finite dimensions}
If a cone $P$ is finite-dimensional then there is unique uniformity making it a topological cone. It corresponds to the unique topology making the enveloping space $EP$ a TVS.

 Then, by Corollary \ref{fd refl. norm}, any finite-dimensional normed cone is reflexive.

  Now, in a finite-dimensional cone, any  closed norm-bounded set is compact, hence, any norm-bounded monotone sequence converges. So a finite-dimensional normed cone is complete.

  The dual of a finite-dimensional normed cone is also finite-dimensional, hence it is also reflexive and complete. Thus, any finite-dimensional cone is a coherent cone.

\subsubsection{Probabilistic coherence spaces}
We briefly recall the notion of {\it probabilistic coherence space} (PCS). For more details see \cite{DanosErhahrdPCS}.

Let $I$ be an at most countable index set.

For any $\alpha\in I$, denote as $e_\alpha$ the sequence indexed by $I$, all whose elements with index other than $\alpha$ are zero, and the element with the index $\alpha$  is $1$.

Let $P$ be a set of real nonnegative sequences indexed by $I$.

The {\it polar} $P^\circ$ of $P$ consists of all real nonnegative sequences $a$ on the index set $I$ such that $\langle a,x\rangle\leq 1$ for any $x\in P$,
 where the pairing of sequences is defined by
 \be\label{PCS pairing}
\langle a,x\rangle=\sum\limits_{i\in I}a_ix_i.
\ee

  A {\it probabilistic coherence space} (PCS) $(P,I)$ is a  pair, where $I$  is an at most countable  index set, and $P$ is a set of real nonnegative sequences on $I$, satisfying the following properties.
\begin{itemize}
  \item $P=P^{\circ\circ}$;
  \item For any $\alpha\in I$ there exist $\lambda, \mu >0$ such that $\lambda e_\alpha\in P$ and $\mu e_\alpha\not\in P$.
  \end{itemize}

The {\it dual PCS} of $(P,I)$ is then the pair $(P^\circ,I)$. (It can be shown that this  is  indeed a PCS.)

If $(P,I)$ is a PCS then the set
$$CP=\{\lambda x|\mbox{ }x\in P, \lambda\in {\bf R}\}$$ is a vector space,
and the subset
$$C_+P=\{\lambda x|\mbox{ }x\in P, \lambda\geq 0\}$$ is a positive cone, making $CP$ a positively generated POVS.

It is easy to see that vector spaces $CP$, $CP^\circ$ form a POVS dual pair under pairing $(\ref{PCS pairing})$.

The space $CP$ is equipped with a norm defined as the {\it Minkowski functional}
$$||x||=\inf\{\lambda>0|\mbox{ } x\in \lambda P\},$$
making $P$  the {\it ``positive unit ball''}, i.e. the set of positive elements with norm less or equal to $1$.

Restriction of the above norm makes $C_+P$ a normed cone, and $(C_+P,C_+P^\circ)$ is then a normed positive dual pair. It is easy to see that the cones are complete.

Thus a PCS $A=(P,I)$ gives rise to the dual pair of coherent cones $(CP,CP^\circ)$, which we will loosely denote with the same letter $A$.
\smallskip

{\bf Remark} Coherent cones coming from PCS have a specific property that  their constituent cones, seen as posets, are {\it lattices}. It can be shown that in the finite-dimensional case this property characterizes PCS in the class of coherent cones completely. In infinite dimensions, however, this property alone apparently is not sufficient. It might be interesting to work out the missing conditions.

Also, the lattice structure determines for PCS preferred bases, and in this sense they are ``commutative'' spaces. In particular, they can be seen as subspaces of commutative algebras, namely, algebras of sequences with pointwise multiplication.
\smallskip

We now discuss morphisms of PCS.

Let $A=(P,I)$, $B=(R,J)$ be PCS.

A {\it PCS morphism} $u:A\to B$ is a double sequence (a matrix) $u=(u_{ij})$ indexed by $I\times  J$, such that for any $x\in P$ the sequence $ux\in R$, where $ux$ is defined by
\be\label{PCS map}
(ux)_j=\sum\limits_{i\in I}u_{ij}x_i.
\ee
(It is implicit in the definition that the series in (\ref{PCS map}) converges for all $j\in J$.)

Now, since $P$ and $R$ are just positive unit balls in the corresponding cones, it is immediate from definition that, for  a PCS map $u:A\to B$, formula (\ref{PCS map}) defines, in fact, a contraction of cones $u:C_+P\to C_+R$. Furthermore, this contraction has the adjoint $u^*:C_+R^\circ\to C_+P^\circ$ given by
$$(u^* \phi)_i=\sum\limits_{j\in J}u_{ij}\phi_j,$$
(note that the  series in the above formula is convergent for all $i$), hence it is continuous.
Thus, a PCS map $u:A\to B$ induces also the map $u:A\to B$ of the corresponding coherent cones.

We are going to show that the converse is true as well: any coherent cone map $u:A\to B$ induces a map of corresponding PCS.

So let $(u,u^*):(C_+P,C_+P^\circ)\to (C_+R,C_+R^\circ)$ be a normed positive dual pair map of norm less or equal to $1$.

Define the matrix $(u_{ij})_{i\in I,j\in J}$ by
\be\label{coherent cone to PCS}
u_{ij}=(ue_i)_j=\langle ue_i,e_j\rangle=\langle e_i,u^*e_j\rangle.
\ee
Let $x\in P$ and $j\in J$. Then
$$(ux)_j=\langle ux,e_j\rangle=\langle x,u^*e_j\rangle=\sum\limits_{i\in I}x_i(u^*e_j)_i=$$
$$=\sum\limits_{i\in I}x_i\langle e_i,u^*e_j\rangle=\sum\limits_{i\in I}x_iu_{ij}.$$

Thus $ux$ is defined by matrix $(u_{ij})$ in the sense of formula (\ref{PCS map}). Also $ux\in R$,  since $x\in  P$, and $u$ is a contraction.

It follows that  the coherent cone map $u:A\to B$ induces the map of corresponding PCS $u:A\to B$ by means of  matrix (\ref{coherent cone to PCS}), and we have the theorem.

\bt The category of PCS and PCS maps is a full subcategory of ${\bf CCones}$. $\Box$
\et

It can be shown that the model of linear logic in ${\bf CCones}$ described in this paper induces the model of \cite{DanosErhahrdPCS} when restricted to PCS. We will not go into these routine details here.

%
%
%
%

\subsubsection{Bounded operators}

A genuinely noncommutative, non-lattice example comes from the space $B(H)$ of bounded operators on a Hilbert space $H$.
  This requires some background, see for example \cite{ConwayOperators}.

      The space $B(H)$ is equipped with the norm
      \be\label{B(H) norm}
       ||L||=\sup\limits_{||v||\leq 1}||Lv||=\sup\{\lambda|\mbox{ }\lambda\mbox{ is in the spectrum of }|A| \},
       \ee
       which makes it a (complex) Banach space.

       The subspace $L^1(H)$ of {\it trace-class} operators consists of operators $M$ with the norm
       \be\label{L1 norm}
       ||M||_1=tr|M|<\infty.
       \ee

       The above norm also makes $L^1(H)$ a complex Banach space, and $B(H)$ becomes the Banach dual of $L^1(H)$ under the Hermitian pairing
      \be\label{operator pairing}
      \langle L,M\rangle = tr(L^\dag M).
      \ee
The space $L^1(H)$ is also a dual; it is the Banach dual of the subspace $K(H)\subseteq B(H)$ of compact operators.

       Now, there are {\it real} subspaces $$B_s(H)\subset B(H),\mbox{ }L^1_s(H)\subset L^1(H)$$
        of {\it self-adjoint} operators, which are real Banach spaces.

Furthermore, for any operator $A\in B(H)$ we have the decomposition $A=\Re A+i\Im A$, where
$$\Re A=\frac{A+A^\dag}{2},\quad \Im A=\frac{A-A^\dag}{2i}$$
and $\Re A,\Im A\in B_s(H)$.
This decomposition provides  isomorphisms
\be\label{operator space complexification}
B(H)\cong B_s(H)\otimes_{{\bf R}}{\bf C},\mbox{ }L^1(H)\cong L^1_s(H)\otimes_{{\bf R}}{\bf C},
\ee
        and  the  spaces $B(H)$, $L^1(H)$ become complexifications of the corresponding real spaces of self-adjoint operators.

        Any norm-bounded linear functional $\phi:L^1_s(H)\to {\bf R}$ extends to a norm-bounded functional $\phi\otimes 1:L(H)\to {\bf C}$. Conversely, any norm-bounded functional $\phi:L^1(H)\to{\bf C}$ sending $L^1_s(H)$ to ${\bf R}$ yields by restriction a norm-bounded functional $L^1_s(H)\to {\bf R}$. It follows that  the real Banach dual of $L^1_s(H)$ is identified with a real subspace of the complex Banach dual of $L^1(H)$, and then it is easy to compute  from (\ref{operator space complexification}) that the real Banach dual of $L^1_s(H)$ is isomorphic to
         $B_s(H)$.
         
         Similarly, it is easy to see that $L_s^1(H)$ is the real Banach dual of the self-adjoint part $K_s(H)$ of $K(H)$.

          The self-adjoint spaces are
       partially ordered, with the positive cones consisting of {\it positive} operators $L$, i.e. those for which $\langle v,Lv\rangle\geq 0$ for all $v\in H$.

       Then it is easy to check that the  cones $B_{s+}(H)$, $L^1_{s+}(H)$ with norms (\ref{B(H) norm}), (\ref{L1 norm}) form a normed cone dual pair. They are also complete in the corresponding weak topologies.
       
%
       Indeed, let $\{X_n\}$ be a norm-bounded monotone non-decreasing sequence in $L^1_{s+}(H)$. Since the space $L^1_s(H)$ is the norm-dual of $K_s(H)$, it follows that the sequence $\{X_n\}$  defines in the limit a norm-bounded positive linear functional $X$ on $K_s(H)$, and $X\in L^1_{s+}(H)$. Now, since for any vector $e\in H$ we have $\langle e,Xe\rangle=\sup\limits_n\langle e,X_ne\rangle=\lim\limits_{n\to\infty}\langle e,X_ne\rangle$ and for any orthonormal basis $\{e_i\}$ of $H$ and $Y\in L^1(H)$ we have $tr(Y)=\sum\langle e_i,Ye_i\rangle$, it follows that  $tr(X)=\sup\limits_ntr(X_i)=\lim\limits_{n\to\infty}tr(X_i)$. Furthermore, all operators $X-X_n$ are positive, hence  $||X-X_n||=tr(X-X_n)$ for all $n$. Thus the sequence $\{X_n\}$ converges to $X$ also in the norm-topology and, consequently, in the $\sigma(L^1_{s+}(H),K_{s+}(H))$-weak topology. Thus the cone $L^1_{s+}(H)$ is complete.
       
       As for the cone $B_{s+}(H)$, it is complete in the $\sigma(B_{s+}(H),L_{s+}^1(H))$-weak topology simply because $B_s(H)$  is the norm-dual of $L^1_s(H)$.

       It might be interesting to try generalizing this example  by considering other normed spaces of self-adjoint (or essentially self-adjoint) operators and pairing (\ref{operator pairing}). A version of such a construction for finite-dimensional Hilbert spaces was hinted by Girard a while ago under the name of {\it quantum coherent spaces} \cite{Girard_Logic_quantic}.

       It is worth noting that the PCS example  also has a ready interpretation as coming from spaces of self-adjoint operators, when the ambient operator algebras are {\it commutative}.

\subsection{Tensor product}
We now show that ${\bf CCones}$ is a $*$-autonomous category, that is, we define tensor and cotensor products and internal homs.

Let $A$, $B$ be coherent cones.

We define the {\it internal homs} normed cone $A\multimap B$ as the set of norm-bounded uniformly continuous  maps from $A$ to $B$  equipped with standard norm (\ref{positive map norm}).

This cone  can also be described as a cone of bilinear functionals.

Let us say that  an ${\bf R}_+$-bilinear  functional $F:A\times  B\to{\bf R}_+$ is
{\it bounded},
  if
   \be\label{norm cotensor positive}
   ||F||=\sup\limits_{||v||,||u||\leq 1)}F(v,u)\leq\infty.
   \ee
   We say that $F$ is {\it uniformly continuous} if for all $v\in A$. $u\in B$, the functionals $F(v,.):B\to {\bf R}_+$, $F(.,u):A\to {\bf R}_+$ are uniformly continuous.

Consider the normed  cone $A\parr B$ of bilinear separately uniformly continuous functionals on $A^*\times  B^*$ bounded in the above  norm. (This is a subcone of the topological cotensor product $A\parr_{top}B$).
\nb
The normed cones $A\multimap B$ and $A^*\parr B$ are isomorphic.
\nbe
{\bf Proof} same as for Note \ref{POVS internal hom}. $\Box$
\smallskip

%
%
%
 \nb\label{cotensor is complete}
  The cones $A\multimap B$, $A^*\parr B$ are  complete, respectively, in $\sigma(A\multimap B,A\otimes_{top}B^*)$-weak and $\sigma(A^*\parr B,A\otimes_{top}B^*)$-weak topology.
 \nbe
 {\bf Proof}
 We prove the statement for $A\multimap B$.

 The topology in question is equivalent to the topology of pointwise convergence on elements of $A\times B^*$, because elements of the form $u\otimes\phi$, where $u\in A$, $\phi\in B^*$, span algebraically the whole $A\otimes B^*$.

Now, if $\{L_n\}$ is a norm-bounded non-decreasing sequence in $A\multimap B$, then for any $u\in A$ the sequence  $\{L_nu\}$ is non-decreasing and bounded in norm by $$||L_nu||\leq\sup\limits_n|||L_n||\cdot||u||$$ (since norms on reflexive cones are monotone). Hence it has a weak limit in $B$. Thus we define
$$Lu=\lim\limits_{n\to\infty} L_nu.$$
Obviously, this is adjointable with the adjoint $L^*$ given by the similar formula
$$L^*\phi=\lim\limits_{n\to\infty} L^*_n\phi,$$
hence it is uniformly continuous by Corollary \ref{adjointable positive}. It is also bounded with norm less or equal to $\sup\limits_n||L_n||$.

So $L\in A\multimap B$. But $L$, seen as a functional on $A\times  B^*$, is precisely the pointwise limit of $\{L_n\}$. $\Box$
\smallskip

Now we equip the topological tensor product of cones $A\otimes_{top}B$ with the norm
$$
||s||=\sup\limits_{f\in A^*\parr B^*,||f||\leq 1}\langle f,s\rangle
$$
(note that the norm is finite).
  We define the {\it tensor product} $A\otimes B$ of coherent cones $A$, $B$ as the completion of the normed cone $A\otimes_{top}B$: $$A\otimes B=\overline{A\otimes_{top}B}.$$

 Finally we topologize the normed cone $A\parr B$  with the topology of pointwise convergence  on $A^*\otimes B^*$.

 It follows from Note \ref{cotensor is complete} and Theorem \ref{completion is complete}
that  $A\parr B$  remains complete in this topology.

Thus we define the {\it cotensor product} $A\parr B$ of coherent cones $A$, $B$ as the normed cone $A\parr B$ above equipped with the $\sigma(A\parr NB,A^*\otimes B^*)$-weak topology.

Finally,  the  {\it internal hom-space} $A\multimap B$ is defined as the normed cone $A\multimap B$ equipped with topology of $A^*\parr B$.

\bt
The category ${\bf CCones}$ is $*$-autonomous, i.e.
there is a natural bijection $Hom(A, B\multimap C)\cong Hom(A\otimes B,C)$. $\Box$
\et
{\bf Proof} follows from Theorem \ref{POVS *-aut} and  Theorem \ref{universal property}. $\Box$

\subsection{Additive structure}
The main use of norms is that they allow defining (non-degenerate) additive structure.

       The {\it product} $P\times  R$ and, respectively, the {\it coproduct} $P\oplus R$  of normed cones $P$ and $R$ are defined as the set-theoretic cartesian product $P\times  R$, equipped respectively with norms \be\label{product norm}
 ||(u,v)||=\max(||u||,||v||),
 \ee
 and
 \be\label{coproduct norm}
 ||(u,v)||=||u||+||v||.
 \ee

 The trivial coherent cone ${\bf 0}$ comes from the trivial cone (or vector space) $\{0\}$.

It is immediate that the above operations  define indeed categorical product and coproduct in ${\bf CCones}$.
Furthermore we have
  $$(A\times  B)^*=A^*\oplus B^*,\mbox{ }(A\oplus B)^*=A^*\times  B^*,$$
 $$A\oplus{\bf 0}\cong A\times {\bf 0}\cong A,$$
 and $A\times  B$ is not isomorphic to $A\oplus B$ unless $A\cong {\bf 0}$ or $B\cong {\bf 0}$.

\section{Exponentials}
Recall that linear logic exponential connectives allow  one to recover the expressive power of intuitionistic logic.

In particular the $!$-modality allows multiple use or waste of  a formula in a proof, thus making a bridge between ``linear'' multiplicative conjunction $\otimes$ and ``intuitionistic'' additive or context-sharing $\&$, which is summarized in the {\it exponential isomorphism}
$$!A\otimes !B\cong !(A\& B).$$

Also, the {\it intuitionistic implication} $A\Rightarrow B$ inside linear logic  defined as
$$A\Rightarrow B=!A\multimap B$$
provides an embedding of intuitionistic logic.

This is reflected on the semantic side.
    A corresponding functor $!$ on a $*$-autonomous category ${\bf C}$ produces a model ${\bf K}$ of intuitionistic logic, whose maps from $A$ to $B$ come from ${\bf C}$-maps from $!A$ to $B$.

  An accurate formulation is in terms of a {\it linear-nonlinear adjunction}. We briefly describe this structure; however, see \cite{Mellies_categorical_semantics} for full details.

 \subsection{Abstract categorical model theory}\label{linear-nonlinear}
  Recall that a linear-nonlinear adjunction consists of the following data:
  \begin{itemize}
    \item a $*$-autonomous category ${\bf C}$ (i.e. a model of multiplicative linear logic);
    \item a  category ${\bf K}$ with a cartesian product $\times $ and a terminal object $T$;
    \item two functors ${\bf R}:{\bf C}\to{\bf K}$ and ${\bf L}:{\bf K}\to{\bf C}$, which are {\it monoidal}, i.e. there are natural transformations
        \be\label{mediating R}
        {\bf R}A\times  {\bf R}B\to {\bf R}(A\otimes B),\mbox{ }T\to{\bf R1},
        \ee
        \be\label{mediating L}
        {\bf L}A\otimes {\bf L}B\to {\bf L}(A\times  B),\mbox{ }{\bf 1}\to{\bf L}T,
        \ee
        and moreover are {\it lax symmetric monoidal}, which means that  the above natural transformations commute in a reasonable way with various combinations of associativity and symmetry morphisms for $\otimes$- and $\times $- monoidal structures;
    \item an {\it adjunction} between ${\bf L}$ and ${\bf R}$, i.e. a natural bijection
    \be\label{adjunction}
    Hom(A,{\bf R}B)\cong Hom({\bf L}A,B),
    \ee
    which is a {\it symmetric monoidal adjunction}.
  \end{itemize}
  The last condition, i.e. that  adjunction (\ref{adjunction}) is symmetric monoidal is equivalent (see \cite{Mellies_categorical_semantics}, Section 5.17) to saying that the monoidal functor ${\bf L}$ is {\it strong}, i.e. that natural transformations (\ref{mediating L}) are isomorphisms.

\bt[\cite{Mellies_categorical_semantics}]\label{model from adjunction}
In  the above setting the endofunctor $!={\bf L}\circ{\bf R}$ equips the category ${\bf C}$ with a model of the linear logic $!$-connective. $\Box$
\et

We now go to the concrete case of coherent cones. As expected, the exponential connectives will be interpreted as spaces of power series, which can be understood as analytic functions and analytic distributions. Using these ingredients, we will construct a category of analytic maps between coherent cones. This will be the second member of a linear-nonlinear adjunction providing us  with a model of the exponential fragment.

\subsection{Symmetric tensor and cotensor}
At first we discuss symmetric (co)tensor algebra in ${\bf CCones}$.

Let $A$ be a coherent cone.

 For any $n\geq 0$ consider the cones $\parr_{top}^nA$ of $n$-linear separately uniformly continuous functionals on $A^*$.

Let  ${{\widehat{\parr}}}_{top}^nA$ be the subcone of {\it symmetric $n$-linear functionals} $a$ in $\parr_{top}^n A$, i.e. those satisfying $$a(x_1,\ldots, x_n)=a(x_{\pi(1)},\ldots,x_{\pi(n)})$$ for any permutation $\pi\in S_n$.

Similarly,  consider the  $n$-th tensor power $\otimes_{top}^nA$.

It carries an action of the permutation group $S_n$, given on generators by $$\sigma x_1\otimes\ldots\otimes x_n=x_{\sigma(1)}\otimes\ldots\otimes x_{\sigma(n)}.$$
Let
the cone ${\widehat\otimes}_{top}^n A$ be the submodule of {\it symmetric tensors} in $\otimes_{top}^n A$ that are fixed by the action of $S_n$.


It is easy to see that $({\widehat\parr}_{top}^nA,{\widehat\otimes}_{top}^n A^*)$ is a cone dual pair. There are different ways to equip its members with norms.

On one hand there are  norms inherited from $\parr^n_{top}A$ and $\otimes_{{alg}}^n A$.

Let us denote  for $a\in {\widehat\parr}_{top}^nA$
 \be\label{norm cotensor^n}
 ||a||_{\parr}=\sup\limits_{||x_1||,\ldots||x_n||\leq1} a(x_1,\ldots,x_n),
 \ee
 and let $||.||_\otimes$ be the  norm on ${\widehat\otimes}_{top}^nA$ defined by duality with ${\widehat\parr}_{top}^nA^*$.

 Let us call these norms {\it old}.

 On the other hand, we define the {\it new} norm for $a\in {\widehat\parr}_{top}^nA$,  by
\be\label{norm cotensor symmetric}
||a||=\sup\limits_{||x||\leq 1}a(x)
\ee
where $a(x)$ is a shorthand for
$$a(x)=a(x,\ldots,x).$$
 The new norm $||.||$ on ${\widehat\otimes}_{top}^nA$ is the dual of the new norm on ${\widehat\parr}_{top}^nA^*$.

It turns out however that the new and the old norms  are equivalent, in the sense that sets bounded in one norm are bounded in the other, and vice versa.
This follows from the following observation.

\nb\label{symmetric tensor  generators}
The cone ${\widehat\otimes}_{top}^nA$ is generated by elements of the form $\otimes^nx$, $x\in A$.
\nbe
{\bf Proof} Obviously, the cone $\widehat\otimes_{top}^nA$ is generated by elements of the form
$$sym_\otimes(x_1\otimes\ldots\otimes x_n)=\frac{1}{n!}\sum\limits_{\sigma\in S_n}a(x_{\sigma(1)},\ldots,x_{\sigma(n)}),\mbox{ }x_1,\ldots,x_n\in A.$$
 But
 \be\label{symmetric tensor expression}
 sym_\otimes(x_1\otimes\ldots\otimes x_n)=\sum\limits_{k=1}^n\sum\limits_{i_1<\ldots<i_k}\otimes^n(x_{i_1}+\ldots+x_{i_k})(-1)^{n-k}.\quad \Box
 \ee

\bc The normed cones $({\widehat\parr}_{top}^nA,||.||)$, and $({\widehat\parr}_{top}^nA,||.||_\parr)$ have the same norm-bounded sets.
\ec
{\bf Proof} Let $a\in {\widehat\parr}_{top}^nA$. Then, obviously,
$$||a||\leq||a||_\parr.$$
On the other hand,  we have from (\ref{symmetric tensor expression})
that
$$||a||_{\parr}\leq\sum\limits_{k=1}^n\frac{n!}{k!(n-k)!}||a||=2^n||a||.\quad \Box$$
\bigskip

Let us denote normed cones $${\widehat\parr}_{old}^nA=({\widehat\parr}_{top}^nA,||.||_\parr),\quad
{\widehat\parr}_{new}^nA=({\widehat\parr}_{top}^nA,||.||),$$
$${\widehat\otimes}_{old}^nA=({\widehat\otimes}_{top}^nA,||.||_\otimes),\quad
{\widehat\otimes}_{new}^nA=({\widehat\otimes}_{top}^nA,||.||).$$

We have that ${\widehat\parr}_{top}^nA$ is a topologically closed subspace of $\parr_{top}^nA$ (in the weak topology), and $\parr_{top}^nA$ is a complete reflexive uniform normed cone. So the unifrom normed cone ${\widehat\parr}_{old}^nA$ is complete. But since the new norm is equivalent to the old, it follows  that ${\widehat\parr}_{new}^nA$ is complete as well.

We define  the {\it $n$-th symmetric tensor power} ${\widehat\otimes}^nA$ as the closure of ${\widehat\otimes}_{new}^n A$ in the $\sigma({\widehat\otimes}_{new}^n A,{\widehat\parr}_{new}^n A^*)$-weak topology.

We define the
{\it $n$-th symmetric cotensor power} ${\widehat\parr}^nA$ as the normed cone ${\widehat\parr}_{new}^nA$ equipped with the $\sigma({\widehat\parr}_{new}^nA,{\widehat\otimes}^nA^*)$-weak topology.

By Theorem \ref{completion is complete} these are indeed coherent cones.

\subsubsection{Interaction with additive operations}
The following easy remarks prepare the exponential isomorphism.

We introduce the following notation. For a $k$-element subset
$$I\subseteq\{1,\ldots,n\}$$
let $i_1,\ldots,i_k$ denote elements of $I$ in the increasing order, and $\overline{i_1},\ldots,\overline{i_n}$ denote elements of the complement of $I$ in the increasing order.
\nb\label{exponential isomorphism preparation}
Let $P$, $Q$ be coherent cones.

The elements $a$ of the cone ${\widehat\parr}^m(P\oplus Q)$ have representation as tuples
$$(a_{0,m},\ldots, a_{k,m-k},\ldots, a_{m,0}),$$
where $a_{k,n}\in ({\widehat\parr}^k P)\parr({\widehat\parr}^nQ)$, such that
$$a((x_1,y_1),\ldots,(x_m,y_m))=\sum\limits_{k+n=m}\sum\limits_{I\subseteq\{1,\ldots,k+n\}, |I|=k}a_{k,n}(x_{i_1},\ldots,x_{i_k},y_{\overline{i_1}},\ldots,y_{\overline{i_n}}).$$
Conversely, any such tuple determines an element of ${\widehat\parr}^m(P\oplus Q)$ by the above formula.
\nbe
{\bf Proof} For $a\in {\widehat\parr}^m(P\oplus Q)$ the coefficients $a_{k,n}$ are given by
$$a_{k,n}(x_1,\ldots,x_k,y_1,\ldots,y_n)=a((x_1,0),\ldots,(x_k,0),(0,y_1),\ldots,(0,y_n)).$$
The other direction is obvious. $\Box$


\subsection{Power series space}
%
%
%
For a normed cone $P$ we denote  the {\it positive unit  ball}, i.e. the set of elements with norm less or equal to $1$, as $B_+(P)$

Let the cone $\widehat?A$ consist  of all sequences
$$a=(a_n),\mbox{ } a_n\in {\widehat\parr}^nA, \mbox{ }n=0,1,\ldots,$$
 such that
 \be\label{exp norm}
||a||=\sup\limits_{x\in B_+(A^*)}\sum a_n(x)<\infty.
\ee
The cone $\widehat?A$ has an interpretation as a space of positive power series $$a=\sum a_n$$ with
$$a_n\in {\widehat\parr}^nA,\mbox{ } n=0,1,\ldots,$$
 which define bounded real analytic functions on $B_+(A)$
 by
\be\label{analytic function}
a(x)=\sum a_n(x).
\ee
 In general, however, these functions are  not continuous (at the positive unit ball boundary)  for any reasonable topology compatible with the norm, as can be observed already in the finite-dimensional case (where the topology is unique).

The second member of the dual pair is the cone
 $\widehat !A$ consisting  of all  sequences
$$x=(x_n),\mbox{ } x_n\in {\widehat\otimes}^nA,\mbox{ } n=1,2\ldots,$$ with only finitely many nonzero elements.

 The cones $\widehat ! A$, $\widehat?(A^*)$ form a cone dual pair under the pairing
\be
\langle a,x\rangle=\sum \langle a_n, x_n\rangle.
\ee
We equip the cone $\widehat!A$ with the norm dual to norm (\ref{exp norm}) of $\widehat?(A^*)$, which makes $(\widehat! A, ?(A^*))$ a normed cone dual pair.

As several times before we will now complete this pair to get a coherent cone.

First, observe that $\widehat?A$ is complete (for the weak topology).

Indeed,  for $a,b\in \widehat?A$ we have
$a\leq b$ iff $a_k\leq b_k$ for all  $k$. Hence if a sequence $\{a_n\}$ in $\widehat?A$ is monotone non-decreasing and norm-bounded, then for any $k$ the coefficient sequence $\{(a_n)_k\}$ is  monotone non-decreasing and norm-bounded, and by completeness of $\widehat\parr^kA$ it has the limit $a_k\in\widehat\parr^kA$. Then, considering the sequence $a=(a_0,a_1,\ldots)$, we observe that for any $x\in B_+(A^*)$ it holds that
$$\sum\limits_k a_k(x)=\sup\limits_N\sum\limits_{k=0}^Na_k(x)=\sup\limits_N\sum\limits_{k=0}^N(\lim_{n\to\infty}a_n)_k(x)=$$
$$=
\sup\limits_N\sum\limits_{k=0}^N(\sup\limits_na_n)_k(x)=\sup\limits_n\sup\limits_N\sum\limits_{k=0}^N(a_n)_k(x)=
\sup\limits_n a_n(x)\leq\sup\limits_n||a_n||<\infty,$$
hence $a\in \widehat?A$, and it is immediate that $a$ is the weak limit of $\{a_n\}$.

So Theorem \ref{completion is complete} guarantees that the following defines coherent cones.

\bd
With notation as above, the cone $!A$ is the completion of $\widehat!A$, and $?A$ is the cone $\widehat?A$ equipped with the topology of pointwise convergence on elements of $!A^*$.
\ed

The  cone $!A$ has then an interpretation as a space of positive  real analytic distributions supported in  $B_+(A)$.

 In particular  $!A$ contains all $\delta$-like distributions $\delta_x$, where  $x\in B_+(A)$, defined on analytic functions by
\be\label{delta definition}
\langle a,\delta_x\rangle=a(x).
\ee
Indeed, any such distribution $\delta_x$ is the limit of the  monotone norm-bounded sequence $\{\sum\limits_{k=0}^n\otimes^kx\}_{n=0}^\infty$.

\subsection{Analytic maps}
Let $P,Q$ be coherent cones.

A map $F:B_+(P)\to Q$ is {\it analytic} if it can be represented as a series
$$F(x)=\sum\limits_nF_n (\otimes^nx)$$
converging in topology of $Q$ for all $x\in B_+(P)$, where $F_n\in {\widehat\otimes}^nP\multimap Q$.

The norm of an analytic map $F:B_+(P)\to Q$ is defined by
$$||F||=\sup\limits_{x\in B_+(P)}||F(x)||.$$

\nb\label{delta}
Let $P$ be a coherent cone.

The map $\delta:B_+(P)\to !P$, sending $x$ to $\delta_x$ is analytic (where $\delta_x$ is defined by (\ref{delta definition})). $\Box$
\nbe

Let $An(P,Q)$ be the set of analytic maps from $B_+(P)$ to $Q$.

\nb There is a norm-preserving bijection
\be\label{adjoint functors}
An(P,Q)\cong !P\multimap Q.
\ee
\nbe
{\bf Proof} Let $F$ be an analytic map as in the formulation.  

The series of coefficients $\sum F_n$ induces a linear map $\widehat!P\to Q$ in the obvious way. This linear map has the adjoint $Q^*\to \widehat?(P^*)$ sending $\phi\in Q^*$ to the function $\phi\circ F$, which is analytic being defined on $x\in B_+(P)$ by the convergent power series $\sum \langle\phi,F_n x\rangle$. By Corollary \ref{adjointable positive} the constructed linear map $\widehat!P\to Q$ is uniformly continuous, and by Theorem \ref{universal property} it extends to a uniformly continuous map $!P\to Q$, i.e. to an element of $!P\multimap Q$.

Conversely, any element $L\in !P\multimap Q$ induces an analytic map on $B_+(P)$ sending $x\in B_+(P)$ to $L\delta_x=\sum L_n \otimes^nx$, where $L_n$ is the restriction of $L$ to $\widehat\otimes ^nP$ (the series on the righthand side being convergent by continuity of $L$).

That this bijection is norm-preserving is immediate. $\Box$

\bl\label{analytic compose}
Analytic maps of norm less or equal to $1$ compose.
\el
{\bf Proof}
Let $P,Q,R$ be coherent cones and $$F:B_+(P)\to Q,\mbox{ }G:B_+(Q)\to R$$
be analytic maps.

 Since $F$ takes $B_+(P)$ to $B_+(Q)$, the composition
 $H=G\circ F$ is well defined as a set-theoretic map.

Define $G_{\leq n}:Q\to R$ by
$$G_{\leq n}(y)=\sum\limits_{k=0}^n G_k(\otimes^k y)$$
and
 $$H_{(\leq n)}=G_{\leq n}\circ F.$$
  Now each $H_{(\leq n)}$ is analytic as the composition of a polynomial and an analytic map. Also the sequence $\{H_{(\leq n)}\}$ is monotone and bounded by $1$ in norm. Hence, under identification (\ref{adjoint functors}) it has a limit $H_{\infty}$ in $!P\multimap R$ with $H_{\infty}=\sup H_{(\leq n)}$. But $\sup H_{(\leq n)}=\sup G_{\leq n}\circ F=G\circ F=H$. So $H\in !P\multimap R$, i.e. $H$ is analytic. $\Box$
\bigskip

Thus  coherent cones and analytic maps form a category ${\bf ExpCCones}$, where morphisms between coherent cones $P$ and $Q$ are analytic maps from $B_+(P)$ to $Q$ of norm less or equal to $1$.
\smallskip

{\bf Remark} The proof of Lemma \ref{analytic compose} is the only place where we explicitly use that our normed cones are complete.

\subsubsection{Exponential isomorphism}
Observe that Cartesian product $\times $ in ${\bf CCones}$ (which is just set-theoretic Cartesian product) extends from objects to analytic maps, providing the category ${\bf ExpCCones}$ with a Cartesian product as well. Similarly, the space $T=\{0\}$ is a terminal object in ${\bf ExpCCones}$ as well as in ${\bf CCones}$.

For modeling the exponential fragment of linear logic, it is crucial how does the Cartesian product in ${\bf ExpCCones}$ interact with the monoidal structure $\otimes$ in ${\bf CCones}$

\nb\label{exponential isomorphism}
In ${\bf CCones}$ there is a natural isomorphism
$$!(P\times Q)\cong !P\otimes !Q.$$
\nbe
{\bf Proof} We establish the dual version $?(P\oplus Q)\cong ?P\parr ?Q$.

Let $P,Q$ be coherent cones. 

By Note \ref{exponential isomorphism preparation} an element
$a\in ?(A\oplus B)$ is represented as a double sequence $(a_{mn})$ with $$a_{mn}\in (\widehat\parr^m P)\parr(\widehat\parr^mQ),$$ such that  the double series
$\sum\limits_{mn}a_{mn}(\otimes^mx,\otimes^ny)$ converges for any  $x\in B_+(P)$, $y\in B_+(Q)$.

On the other hand (using identification (\ref{adjoint functors})), an element $a\in ?P\parr?Q$ is represented as a double sequence $(a_{mn})$, such that for any $x\in B_+(P)$, $y\in B_+(Q)$ the repeated series
$\sum\limits_{m}\sum\limits_na_{mn}(\otimes^mx,\otimes^ny)$ converges. But since all terms are nonnegative,   convergence of the repeated series is equivalent to convergence of the double series (the sum of a nonnegative series is the supremum of its partial sums). $\Box$
\bigskip

We now define functors connecting ${\bf CCones}$ and ${\bf ExpCCones}$ yielding a linear-nonlinear adjunction.

\subsection{The adjunction}
There is the obvious functor
 $${\bf R}: {\bf CCones}\to{\bf ExpCCones}$$
  sending  coherent cones and uniformly continuous contractions to themselves (since a linear map is obviously an analytic map).

Now, if $$F:B_+(P)\to Q$$ is analytic  of norm less or equal than $1$,  we get another analytic map
$$\delta\circ F:B_+(P)\to !Q$$
(by Lemma \ref{analytic compose}), which corresponds, under identification (\ref{adjoint functors}), to an element $${\bf L}F\in !P\multimap !Q.$$
 In particular ${\bf L}F$ represents a uniformly continuous ${\bf R}_+$-linear map from $!P$ to $!Q$, which is a contraction, hence a morphism in ${\bf CCones}$.

The corresponding analytic map $B_+(P)\to !Q$ (i.e. $\delta\circ F$) sends $x\in B_+(P)$ to $\delta_{F(x)}$, and it is immediate  that the assignment $F\mapsto {\bf L}F$ is functorial. We extend it to a functor by putting ${\bf L}P=!P$ for $P\in {\bf ExpCCones}$.

Correspondence (\ref{adjoint functors}) readily extends to the following.
\nb
There is a natural correspondence
\be\label{adjunction concrete}
Hom({\bf L}P,Q)\cong Hom(R,{\bf R}Q),
\ee
where the lefthand side represents ${\bf CCones}$-morphisms, and the righthand side represents ${\bf ExpCCones}$-morphisms. $\Box$
\nbe

The {\it exponential functor} $!:{\bf CCones}\to{\bf CCones}$ is defined as the composition $!={\bf L}\circ {\bf R}$. Explicitly, the functor $!$ sends coherent cone $A$ to $!A$ and continuous contraction $S:A\to B$ to the element $!S\in !A\multimap !B$, which, under identification (\ref{adjoint functors}), represents the analytic map $x\mapsto\delta_{Sx}$.

Now, if we make sure that correspondence (\ref{adjunction concrete}) is indeed a linear-nonlinear adjunction, then, by Theorem \ref{model from adjunction}, we get indeed a model of full propositional linear logic.
It is sufficient to establish the following.

\nb
Functors ${\bf L}$, ${\bf R}$ are lax symmetric monoidal, moreover ${\bf L}$ is strong.
\nbe
{\bf Proof} All statements about ${\bf L}$ are obvious after Note \ref{exponential isomorphism}.
It remains to observe that ${\bf R}$ is also (lax symmetric) monoidal.
The map $${\bf R}A\times  {\bf R}B\to {\bf R}(A\otimes B)$$ is given explicitly as the function
$$(x,y)\mapsto x\otimes y. \Box$$

\bc
The above defined functor $!:{\bf CCones}\to{\bf CCones}$ is a model of linear logic $!$-modality. $\Box$
\ec

\subsection{Interpretation of exponentials}
  We now  briefly describe the interpretation of basic principles of the exponential fragment.

 The functor $!$ has been discussed above.
    The action of the functor $?$ on morphisms is as follows. For the morphism $$L:A\to B,$$ the morphism $$?L:?A\to?B$$ sends a function $f\in ?A$ to the function $$?Lf=f\circ L^*,$$ which belongs to $?B$.

For any coherent cone $A$ the object $?A$ is naturally a monoid in ${\bf CCones}$.

We have the {\it monoid unit}
 $$u_A:{\bf 1}\to ?A,$$
 sending a number to the corresponding constant function, and the {\it multiplication}
   $$m_A:?A\parr?A\to ?A,$$  given by  the {\it diagonalization} map which sends
the function $f$ of two variables to the function
$$x\mapsto f(x,x).$$
By duality, $!A$ is a {\it comonoid} in ${\bf CCones}$ with the corresponding dual comonoidal maps.

  We also have the important  monadic maps
  $$\eta_A:A\to ?A, \mbox{ }\mu_A:??A\to ?A,$$ used to interpret principles of dereliction and digging.

  The map $\eta_A$  sends an element of $A$ to the corresponding linear function on $A^*$. The map $\mu_A$ is defined by
  $$
  (\mu_Af)(x)= f(\delta_x).
  $$

\section{Conclusion and further work}
We proposed a nondegenerate model of full linear logic, which apparently fits in the familiar tradition of interpreting linear logic in the language of vector spaces, but, arguably, is free from various drawbacks of its predecessors. In particular the model is invariant in the sense that we do not refer to any bases. It also encompasses probabilistic coherence spaces as a ``commutative'' submodel.

There are also genuinely noncommutative objects, especially coming from spaces of self-adjoint operators on Hilbert spaces. The latter look as a variation of Girard's {\it quantum coherence spaces} \cite{Girard_Logic_quantic} (see Section \ref{coherent cone examples}). They might be interesting to study, in particular, it might be interesting to give directly an explicit description of  linear logic connectives specialized to such spaces.

A more important problem, as it seems, is to clarify relations with categories of stable and measurable cone maps of \cite{EhrhardPaganiTasson}. This is left for future research.

 \end{document}